\documentclass[letterpaper,twocolumn,accepted=2026-02-20]{quantumarticle}
\pdfoutput=1

\usepackage{ifpdf}
\usepackage{braket}
\usepackage{amsmath} 
\usepackage{amssymb}
\usepackage{amsfonts}
\usepackage{amsthm}
\usepackage{braket}
\usepackage{xcolor}
\usepackage{hyperref}
\usepackage{graphicx}
\usepackage{dcolumn}
\usepackage{bm}
\usepackage[normalem]{ulem}
\usepackage{verbatim}
\usepackage{orcidlink}
\usepackage[numbers,sort&compress]{natbib} 

\newcommand{\la}{\langle}
\newcommand{\ra}{\rangle}
\newcommand{\tr}{\text{Tr}}

\newcommand{\im}{\text{Im}}
\newcommand{\re}{\text{Re}}

\usepackage{mathtools}

\begin{document}

\title{Utility-Scale Quantum State Preparation: \newline
Classical Training using Pauli Path Simulation
}

\author{Cheng-Ju Lin\,\orcidlink{0000-0001-7898-0211}}
    \affiliation{BlueQubit Inc., San Francisco, California 94105, USA}
    \affiliation{Joint Center for Quantum Information and Computer Science, NIST/University of Maryland, College Park, Maryland 20742, USA}
    \affiliation{Joint Quantum Institute, NIST/University of Maryland, College Park, Maryland 20742, USA}
\author{Hrant Gharibyan\,\orcidlink{0000-0001-5910-7943}}
\affiliation{BlueQubit Inc., San Francisco, California 94105, USA}
\author{Vincent P. Su}
\affiliation{BlueQubit Inc., San Francisco, California 94105, USA}

             
\begin{abstract}
We use Pauli Path simulation to variationally obtain parametrized circuits for preparing ground states of various quantum many-body Hamiltonians. These include the quantum Ising model in one dimension, in two dimensions on square and heavy-hex lattices, and the Kitaev honeycomb model, all at system sizes of one hundred qubits or more --- sizes at which generic quantum circuits are beyond the reach of exact state-vector simulation--- thereby reaching utility scale. We benchmark the Pauli Path simulation results against exact ground-state energies when available, and against density-matrix renormalization group calculations otherwise, finding strong agreement. To further assess the quality of the variational states, we evaluate the magnetization in the $x$ and $z$ directions for the quantum Ising models and compute the topological entanglement entropy for the Kitaev honeycomb model. Finally, we prepare approximate ground states of the Kitaev honeycomb model with $48$ qubits, in both the gapped and gapless regimes, on Quantinuum’s System Model H2 quantum computer using parametrized circuits obtained from Pauli Path simulation. We achieve a relative energy error of approximately $5\%$ without error mitigation and demonstrate the braiding of Abelian anyons on the quantum device beyond fixed-point models.
\end{abstract}

      
\hypersetup{
  colorlinks=true,
  linkcolor=quantumviolet,
  citecolor=quantumviolet,
  urlcolor=quantumviolet
}                              
\maketitle

\section{Introduction}\label{sec: intro}
The development of quantum computers has ushered in the era of noisy intermediate-scale quantum devices~\cite{preskillQuantum2018} --- machines capable of executing quantum circuits with hundreds of qubits, albeit in the presence of noise. Among potential applications, quantum simulation is arguably the most promising avenue for these devices to demonstrate practical advantages over classical methods~\cite{kimEvidence2023,kingBeyondclassical2025,haghshenasDigital2025}. Such applications include simulating the dynamics of quantum many-body systems and probing the ground-state or thermal-state properties of a given quantum Hamiltonian.

Preparing a ground state on a quantum device, even approximately, is often a crucial prerequisite for these tasks. For instance, in quench dynamics~\cite{kingBeyondclassical2025} or scattering simulations~\cite{farrellQuantum2024,farrellDigital2025}, the initial state is typically the ground state of an interacting Hamiltonian. Likewise, to investigate the properties of a ground state or its dynamical responses, high-quality ground-state preparation is essential.

Variational quantum algorithms (VQAs)~\cite{peruzzoVariational2014,mccleanTheory2016,tillyVariational2022} provide a widely used framework for ground-state preparation and are particularly well suited to near-term quantum devices. In this approach, parameterized gates within structured circuits define a family of variational wavefunctions. A quantum device is then used to estimate the energy expectation value of a given variational wavefunction, and often its gradient, with the results passed to a classical computer that updates the circuit parameters according to an optimization algorithm. Despite their conceptual simplicity, implementing VQAs on present-day hardware remains resource-intensive and costly.

A natural strategy to mitigate these challenges is to offload as much computation as possible to classical machines~\cite{huangTensornetworkassisted2023,lerchEfficient2024,mckeeverAdiabatic2024,liDual2025}. In particular, classical simulations can be used to train circuit parameters, with quantum hardware subsequently employed only for tasks such as simulating quantum dynamics from the classically trained quantum state. 
This hybrid approach can substantially reduce runtime and resource demands on quantum devices. 
However, conventional classical simulation methods face limitations: exact state-vector approaches are restricted to small system sizes, while tensor-network methods are typically effective only for shallow circuits or systems with one-dimensional geometries, with performance constrained by entanglement growth.

Complementing these methods, in this work we employ the recently developed coefficient-truncation Pauli Path simulation (PPS)---also known as sparse Pauli dynamics or Pauli propagation~\cite{begusicRealtime2025,begusicFast2024,schusterPolynomialtime2024,gharibyanPractical2025,rudolphPauli2025}---to simulate VQAs for ground-state preparation, targeting system sizes of around one hundred qubits.  At this scale, generic quantum circuits are beyond the reach of exact state-vector simulation, dubbed the utility scale.
However, this does not preclude the possibility that other classical methods, such as tensor network methods or Monte Carlo techniques, may still be able to simulate certain problems at comparable system sizes, provided those problems exhibit favorable structure for such methods.
Coefficient truncation is one of several truncation schemes within the PPS framework~\cite{aharonovPolynomialTime2023,rudolphPauli2025,fontanaClassical,angrisaniSimulating2025,angrisaniClassically2025}. In essence, PPS estimates the expectation value of a quantum circuit by evolving observables in the Heisenberg picture, expanding them in the Pauli-string basis, and discarding terms with coefficients below a chosen threshold to reduce computational cost. This approach has already been successfully applied to quantum many-body dynamics at the utility scale~\cite{begusicRealtime2025,begusicFast2024}, with benchmarking against quantum hardware~\cite{kimEvidence2023} demonstrating comparable precision in certain parameter regimes.

The PPS method has been shown to be effective for simulating noisy quantum circuits at sufficiently high noise rates across various truncation schemes~\cite{aharonovPolynomialTime2023,schusterPolynomialtime2024,fontanaClassical,angrisaniSimulating2025}. 
On the other hand, early numerical studies indicate that weight- and path-based truncation schemes can yield high-quality results for VQAs in condensed matter systems of a small system size~\cite{angrisaniClassically2025}, as well as for quantum machine learning tasks~\cite{bermejoQuantum2024}. 
Motivated by these findings, we investigate the performance of the coefficient-truncation PPS in simulating noiseless VQAs for ground-state preparation, emphasizing system sizes at the utility scale.

To this end, we apply PPS to simulate VQAs for a variety of quantum many-body Hamiltonians, aiming to obtain parametrized circuits that prepare approximate ground states of systems with over one hundred qubits. Specifically, we consider the quantum Ising model in one dimension (1D), in two dimensions (2D) on both square and heavy-hex lattices, and the Kitaev honeycomb model~\cite{Kitaev2005}. Our parametrized circuits are based on the Hamiltonian variational ansatz~\cite{weckerProgress2015,hoEfficient2019,wiersemaExploring2020,parkEfficient2024,parkHamiltonian2024}, with parameter updates performed using a combination of simultaneous perturbation stochastic approximation~\cite{spallIntroduction2003} and adaptive moment estimation~\cite{kingmaAdam2017}.

For models whose ground-state energies can be determined exactly, such as the 1D transverse-field Ising model and the Kitaev honeycomb model, we benchmark PPS results against exact values. 
For models without exact ground-state solutions, we use the density-matrix renormalization group (DMRG)~\cite{whiteDensity1992} as a reference, using the ITensor library~\cite{itensor}. 
Across all cases, we observe high relative energy accuracy, particularly when the ground states are gapped. 
Even in gapless or near-critical regimes, the PPS variational state maintains strong performance. 
Notably, for the 2D Ising model on a heavy-hex lattice, the PPS variational state achieves variational energies lower than those obtained with DMRG across a range of Hamiltonian parameters, both using comparable and modest computational resources. 
Beyond energies, we also compare magnetization in the $x$ and $z$ directions against benchmark results. 
For the Kitaev honeycomb model, we further assess the variational ground state by computing its topological entanglement entropy~\cite{kitaevTopological2006,levinDetecting2006}.

Although the models studied here can be solved exactly or efficiently treated with classical methods, preparing their ground states on a quantum device is not always straightforward. 
The parametrized circuits produced by PPS provide explicit and implementable protocols for this task. 
To demonstrate this, we prepare approximate ground states of the Kitaev honeycomb model in both the gapped and gapless regimes for a 48-qubit system on Quantinuum’s System Model H2 quantum computer~\cite{quantinuum2025}, using the parametrized circuits obtained from PPS. 
Without error mitigation, we achieve relative energy errors of approximately $5\%$. 
Furthermore, we demonstrate anyon braiding and extract the corresponding braiding statistics on these approximate ground states, thereby confirming the topological properties of the quantum states prepared on the Quantinuum machine.
Notably, previous demonstrations of anyon braiding on quantum devices have focused mainly on fixed-point or zero-correlation-length models~\cite{satzingerRealizing2021,niuDemonstrating2024,iqbalTopological2024,iqbalNonAbelian2024a,minevRealizing2025}, whereas our results extend braiding demonstrations beyond the fixed-point regime.

To summarize, in this work we benchmark PPS parametrized circuits for ground-state preparation on prototypical models with well-understood physics, enabling a clear evaluation of the method’s capabilities through comparisons with exact and DMRG results. In certain models and parameter regimes, PPS variational circuits even outperform DMRG, underscoring their potential as a quantum-inspired classical numerical method for ground-state problems in condensed matter and quantum chemistry. 
The method’s suitability for quantum devices is further demonstrated by our preparation of the Kitaev honeycomb ground state and the subsequent braiding of anyons on the quantum hardware. 
We expect these PPS variational states can also serve as high-quality starting points for quantum dynamics simulations or as warm-start parameters for additional variational optimization on a quantum device. 
Such pre-trained parameters substantially reduce resource demands and runtime, positioning PPS variational circuits as an effective bridge between classical simulation and quantum execution.

The rest of the paper is organized as follows. In Sec.~\ref{sec: sparse Pauli}, we provide an overview of the coefficient-truncation PPS method and its application to variational quantum algorithms. We then apply our approach to the quantum Ising model in one and two dimensions, with results presented in Sec.~\ref{sec:1d Ising} and Sec.~\ref{sec: ising 2D}, respectively, using ground-state energies and magnetization as benchmarks. In Sec.~\ref{sec: kitaev honeycomb}, we present results for the Kitaev honeycomb model, benchmarking both energy and topological entanglement entropy. Section~\ref{sec: quantum validation} demonstrates the implementation of parametrized circuits on quantum hardware for a 48-qubit Kitaev honeycomb system in both gapped and gapless regimes. The braiding demonstration on the quantum device highlights the topological nature of the prepared quantum states beyond fixed-point models. Finally, in Sec.~\ref{sec: discussion}, we summarize our findings, discuss their implications, and outline directions for future research.

\section{Pauli path-simulated variational quantum algorithm}\label{sec: sparse Pauli}

\begin{figure}
    \centering
    {\includegraphics[width=0.48\textwidth]{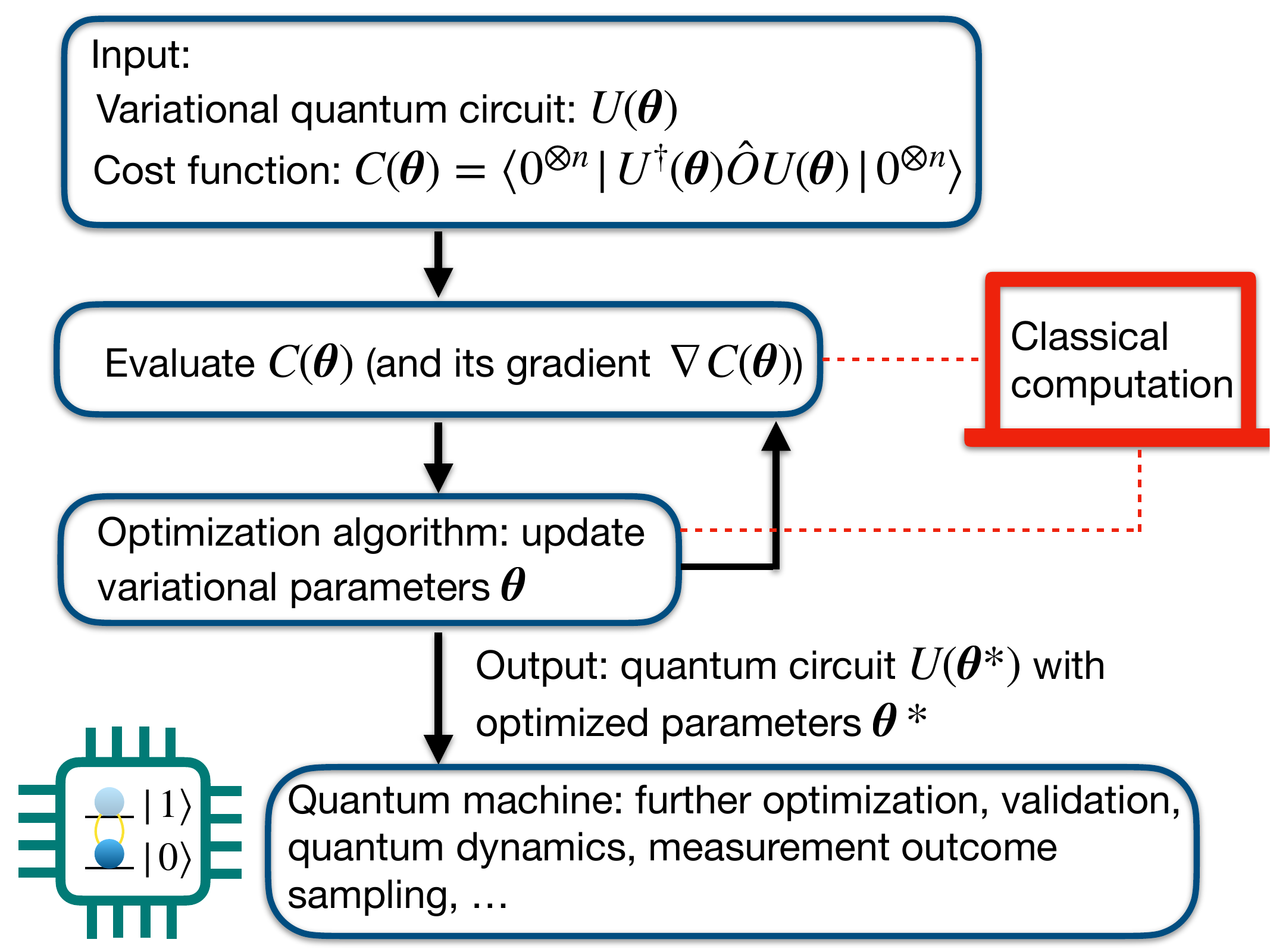}
    }
    \caption{\label{fig:spd vqe} Flowchart of the variational quantum algorithm framework. In the standard approach, the cost function and its gradient are evaluated on a quantum device. In this work, we instead use the coefficient-based Pauli Path simulation to perform these evaluations, allowing us to optimize the parameters prior to executing tasks on the quantum hardware.}
    
\end{figure}

For the task of ground-state preparation, VQA is a widely used framework designed to be compatible with near-term devices, as illustrated schematically in Fig.~\ref{fig:spd vqe}. Given a parameterized circuit $U(\boldsymbol{\theta})$, where the variational parameters are $\boldsymbol{\theta} = (\theta_1,\ldots, \theta_p)$, the objective of VQA is to minimize the cost function
\begin{equation}\label{eqn:cost function Ham}
C(\boldsymbol{\theta}) = \langle 0^{\otimes N} | U^{\dagger}(\boldsymbol{\theta}) \hat{O} U(\boldsymbol{\theta}) | 0^{\otimes N} \rangle,
\end{equation}
for some observable $\hat{O}$. For the ground-state problem, one sets $\hat{O} = \hat{H}$, the system Hamiltonian, in which case the cost function corresponds to the variational energy, and the state $U(\boldsymbol{\theta}) | 0^{\otimes N} \rangle$ corresponds to the variational ground state.

In a standard VQA setup, a quantum device is used to estimate the cost function $C(\boldsymbol{\theta})$ and, often, its gradient $\nabla C(\boldsymbol{\theta})$. These estimates are then passed to a classical optimizer, which updates the variational parameters. 
In this work, however, instead of estimating the cost function and its gradient on quantum hardware, we employ coefficient-truncation PPS (also known as the sparse Pauli method)~\cite{begusicRealtime2025,begusicFast2024,schusterPolynomialtime2024,gharibyanPractical2025} --- a variant of the framework of Pauli Path simulation that incorporates various truncation schemes~\cite{aharonovPolynomialTime2023,rudolphPauli2025,fontanaClassical,angrisaniSimulating2025,angrisaniClassically2025}.

The coefficient-truncation PPS proceeds as follows.
Rather than evolving the state $U(\boldsymbol{\theta})|0^{\otimes N}\ra$ in the Schr\"{o}dinger picture, we evolve the observable in the Heisenberg picture $\hat{O}'=U(\boldsymbol{\theta})^{\dagger}\hat{O}U(\boldsymbol{\theta})$.
Using Pauli strings $\hat{P} \in \{I, X, Y, Z\}^{\otimes N}$ as a basis, any operator can be expanded as $\hat{O}=\sum_{P}a_p \hat{P}$ with coefficients $a_P$. A full representation of $\hat{O}'$ requires $4^N$ terms, scaling exponentially with the number of qubits.
To reduce this cost, PPS truncates Pauli strings with coefficients below a chosen threshold $\delta_c$, yielding an approximate representation sufficient to estimate expectation values.

In particular, the evolution and truncation are performed iteratively gate by gate.
Suppose the quantum circuit $U(\boldsymbol{\theta})$ is composed of a collection of quantum gates $V = \exp(-i\theta \hat{\sigma}/2)$ where $\hat{\sigma}$ is a Pauli string.
Given an operator $\hat{O}=\sum_{P}a_p \hat{P}$,  conjugation by $V$ yields
\begin{align}
    V^{\dagger}\hat{O}V =\sum_{\hat{P} \in P_C}a_P \hat{P} + \sum_{\hat{P} \in P_A} (a_P\cos(\theta)\hat{P} + i\sin(\theta) \hat{\sigma}\cdot \hat{P} ) 
    ~, 
\end{align}
where $P_C$ and $P_A$ denote the sets of Pauli strings that commute and anti-commutes with $\sigma$, respectively.
One then only retains the Pauli strings with the new coefficients $|a_P'| > \delta_{c}$, where $a_P'$ is the coefficients of $V^{\dagger}\hat{O}V = \sum_{P}a_p' \hat{P}$ in the Pauli-string basis.
Repeating this procedure for each gate in the circuit yields an approximate Heisenberg-evolved operator $\hat{O}^{\prime} \approx U^{\dagger} \hat{O} U$,  from which the expectation value $\la 0^{\otimes N}|U^{\dagger}\hat{O}U|0^{\otimes N}\ra \approx \la 0^{\otimes N}|\hat{O}^{\prime}|0^{\otimes N}\ra$ can be approximated.
We expect this approximation to perform well when the depth of the quantum circuit is low, or when the gates of the quantum circuit are near-Clifford, namely, when $\theta$ is close to $0$ or $\pi/2$.

The PPS-estimated cost function and its gradient are then used to update the variational parameters. A typical gradient-based algorithm such as gradient descent requires evaluating the full gradient vector. When computed via finite differences, it requires $2p$ cost-function evaluations for $p$ variational parameters, leading to significant overhead for large $p$. 
To alleviate this, we adopt a variation of the simultaneous perturbation stochastic approximation (SPSA) algorithm combined with adaptive moment estimation (ADAM). 
In brief, SPSA estimates the gradient vector by sampling a random direction in parameter space, requiring only two cost-function evaluations per update, independent of the number of variational parameters. 
ADAM, widely used in machine learning and artificial intelligence, is an adaptive optimization method in which each variational parameter has an individual, dynamically adjusted learning rate determined by the history of updates, encoded through “momentum” and “velocity” vectors. 
We discuss further details on these methods and their implementation in Appendix~\ref{app: SPSA and ADAM}.

\section{Quantum Ising model in 1D}\label{sec:1d Ising}
We begin by applying our method to the 1D quantum Ising model as a test case, whose Hamiltonian is 
\begin{equation}\label{eqn: 1d Ising}
\hat{H}=-\sum_{j=1}^{N}Z_jZ_{j+1}-g_x\sum_{j=1}^NX_j-g_z\sum_{j=1}^NZ_j~,
\end{equation}
where periodic boundary conditions (PBC) are imposed, $j+N \equiv j$, and $N$ denotes the number of qubits.
When $g_z = 0$ (the transverse-field Ising model), the model exhibits a quantum phase transition at $g_x = g_c =1$. For $g_x < 1$, the $\mathbf{Z}_2$ symmetry generated by $\prod_{j=1}^N X_j$ is spontaneously broken, corresponding to the ferromagnetic phase. For $g_x > 1$, the system is in the paramagnetic phase. 
For any nonzero $g_z$, the low-energy degrees of freedom of the model is described by the Ising field theory~\cite{delfinoSpinspin1995,fonsecaIsing2001a}, making this model a valuable setting for studying scattering phenomena in that theory~\cite{vandammeRealtime2021,milstedCollisions2022,jhaRealTime2024,farrellDigital2025,bennewitzSimulating2025}. Moreover, the model has also become a popular playground for exploring various quantum thermalization phenomena~\cite{banulsStrong2011,kormosRealtime2017,linQuasiparticle2017,PhysRevB.96.214301}.

To approximately prepare the ground state of the Hamiltonian, we employ the parametrized circuit with symmetry-breaking capability~\cite{parkEfficient2024} as follows. 
Define 
\begin{equation}
    U(\alpha,\beta,\gamma)=u_{Z}(\alpha)u_{X}(\beta)u_{ZZ}(\gamma)~,
\end{equation}
where 
\begin{align}
    u_{Z}(\alpha) &= \prod_{j=1}^{N}\exp(-i\frac{\alpha}{2}Z_j) \notag \\
    u_{X}(\beta) &= \prod_{j=1}^{N}\exp(-i\frac{\beta}{2}X_j) \notag \\
    u_{ZZ}(\gamma) &= \prod_{j=1}^{N}\exp(-i\frac{\gamma}{2}Z_jZ_{j+1})~.
\end{align}
The variational wavefunction with a parametrized circuit of repetition $\ell$ is constructed as
\begin{equation}\label{eqn: ising hva wavefunction}
|\psi(\boldsymbol{\alpha},\boldsymbol{\beta},\boldsymbol{\gamma})\rangle
= U(\alpha_\ell,\beta_\ell,\gamma_\ell)\cdots U(\alpha_1,\beta_1,\gamma_1)|+\rangle^{\otimes N}~,
\end{equation}
where $|+\rangle$ is the $+1$ eigenstate in the $X$-basis $X|+\rangle = |+\rangle$, and the circuit parameters are grouped as $\boldsymbol{\alpha}=(\alpha_1,\dots,\alpha_\ell)$, with analogous definitions for $\boldsymbol{\beta}$ and $\boldsymbol{\gamma}$.

In the VQA framework, the cost function and its gradient are typically evaluated on a quantum device. As described in Sec.~\ref{sec: sparse Pauli}, we instead approximate them classically using PPS, with a default truncation threshold of $\delta_c = 10^{-3}$. In practice, dynamically adjusting $\delta_c$ during optimization is advantageous. For example, in the early stages, when gradient magnitudes are relatively large, a higher threshold yields sufficiently accurate estimates while substantially reducing computational cost. Furthermore, reusing circuit parameters obtained at nearby model parameters as a warm start can significantly accelerate optimization.

\begin{figure}[!htbp]
    \centering
    {\includegraphics[width=0.48\textwidth]{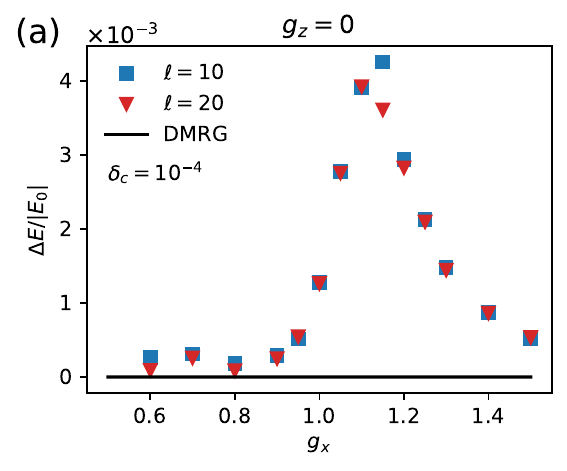}\\
    \includegraphics[width=0.235\textwidth]{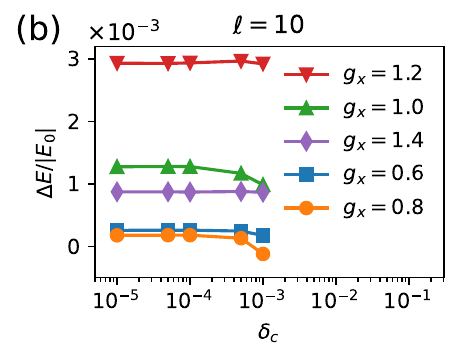}
    \includegraphics[width=0.235\textwidth]{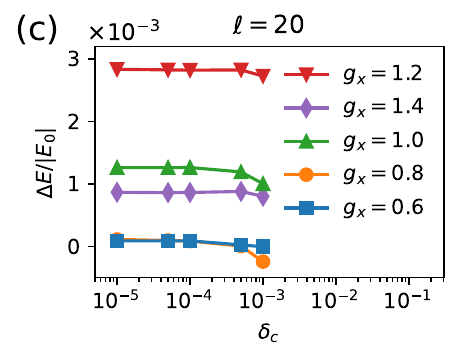}
    }
    \caption{(a) The relative energy error $\Delta E / |E_0|$ for the 1D quantum Ising model with PBC and $N=100$ at $g_z=0$, where $\Delta E = \langle H \rangle - E_0$. The variational energies $\langle H \rangle$ are re-evaluated at $\delta_c = 10^{-4}$, with circuit repetitions $\ell = 10$ and $\ell = 20$. The relative errors remain below $0.5\%$.
    (b)(c) Dependence of the relative error on the truncation threshold $\delta_c$ for repetitions $\ell = 10$ and $\ell = 20$, respectively. The results indicate indicate that $\delta_c = 10^{-4}$ already yields accurate energies.}

    \label{fig:1dsing_gz0}
\end{figure}

The optimized parameters obtained with a looser threshold are subsequently used to compute the final variational energy $\langle \hat{H} \rangle$ at a more stringent value (e.g., $\delta_c = 10^{-4}$) to improve accuracy. In principle, this final result may also be benchmarked against direct measurements on a quantum device, which corresponds to evaluating the cost function at $\delta_c = 0$, modulo hardware noise. To compute gradients efficiently, we employ a modified SPSA algorithm in combination with ADAM optimization, as detailed in Appendix~\ref{app: SPSA and ADAM}.

\subsection{Transverse-field quantum Ising model}\label{subsec:ising_1d gz = 0}

We begin by examining results in the regime where the longitudinal field $g_z = 0$, for various values of the transverse field $g_x$, using a system of $N = 100$ qubits.
In this case, Eq.~(\ref{eqn: 1d Ising}) is exactly solvable, and the ground state energy $E_0$ can be obtained analytically~\cite{LIEB1961407}. For comparison, we also compute the ground state and its energy $E_{\text{DMRG}}$ numerically using DMRG.

Figure~\ref{fig:1dsing_gz0}~(a) shows the relative error $\Delta E / |E_0|$, where $\Delta E = \langle \hat{H} \rangle - E_0$, calculated from the variational energy $\langle \hat{H} \rangle$ [Eq.(\ref{eqn:cost function Ham})] at various values of $g_x$, for repetitions $\ell = 10$ and $\ell = 20$. While the optimization is typically performed using PPS with truncation threshold $\delta_c = 10^{-3}$, the variational energies plotted here are re-evaluated at a more stringent threshold $\delta_c = 10^{-4}$. As shown in the figure, the relative errors remain below $0.5\%$, with slightly larger deviations appearing near the critical point $g_c = 1$.

Figs.~\ref{fig:1dsing_gz0}(b) and (c) further illustrate the dependence of the relative error on $\delta_c$ for selected values of $g_x$. We find that $\delta_c = 10^{-4}$ already provides sufficiently accurate energy estimations, as decreasing $\delta_c$ further produces negligible changes in the estimated energies.

Interestingly, we also observe cases where truncation leads to trial energies that fall below the true ground-state energy [e.g., at $g_x = 0.8$ in Figs.~\ref{fig:1dsing_gz0}(b) and (c)]. This artifact arises because the truncation scheme does not preserve the unitarity of the circuit. Nevertheless, evaluating the energy at a more stringent threshold alleviates this issue, and the circuit parameters obtained from a looser truncation, even when yielding an estimated energy lower than the true ground state energy, still serve as excellent solutions.

\begin{figure*}
    \centering
    {\includegraphics[width=0.48\textwidth]{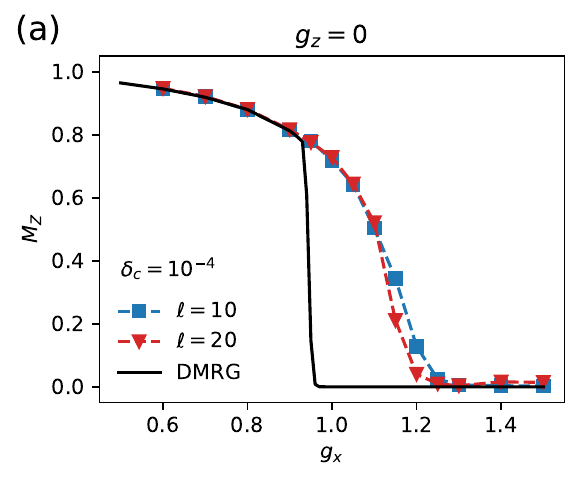}
    \includegraphics[width=0.48\textwidth]{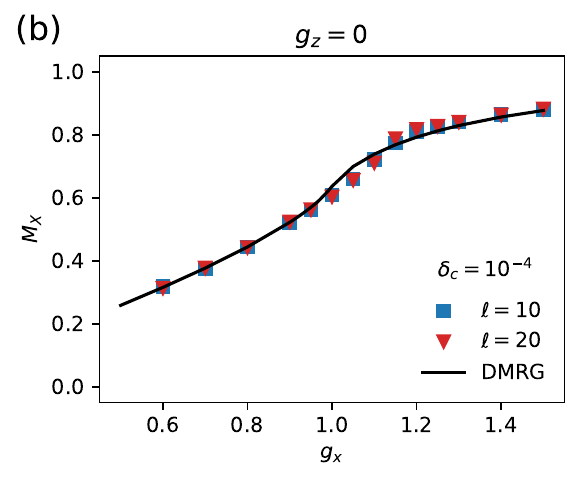}
    }
    \caption{\label{fig:1dising_gz0_Mz_Mx}
    For the the 1D quantum Ising model with PBC and $N=100$ at $g_z=0$, (a) $M_z$ and (b) $M_x$ for the variational wavefunction with circuit repetitions $\ell =10$ and $\ell =20$, evaluated at $\delta_c=10^{-4}$. The results show good agreement with the DMRG data away from the critical point $g_c =1$. Near the critical point, we expect the results can be improved by increasing the circuit repetition.}

\end{figure*}

Another valuable benchmark involves evaluating the expectation values of physical observables.
To this end, we compute the magnetization in the $z$ and $x$ directions,
\begin{align}
    M_z := \frac{1}{N}\sum_{j=1}^{N}\la Z_j \ra~,~ &~M_x := \frac{1}{N}\sum_{j=1}^{N}\la X_j \ra~, 
\end{align} evaluated on our variational wavefunction at $\delta_c=10^{-4}$.
These quantities are shown in Figs.~\ref{fig:1dising_gz0_Mz_Mx}(a) and (b), where they are compared against reference results obtained via DMRG.

For values of $g_x$ away from the critical point, both $M_z$ and $M_x$ exhibit excellent agreement with the DMRG benchmarks, confirming the reliability of our variational ansatz in those regimes. Near the critical point, however, the ansatz tends to overestimate $M_z$, and $M_x$ also shows slight deviations from the DMRG result. 
Increasing the repetition from $\ell=10$ to $\ell=20$, we observe the relative energy error and the magnetization benchmarks improved slightly. 
This suggest that greater repetition $\ell$ could improve accuracy in the critical regime.
At the same time, the performance of PPS may degrade as circuit depth increases due to a more complex optimization landscape and greater simulation requirements. This trade-off highlights a potential practical advantage of executing VQAs on quantum hardware near criticality, where deeper circuits are needed to capture the strong correlations.

In fact, we observe a common trend across the models examined in this work: our PPS results tend to perform less well near the critical point or in the gapless regime. We suspect that this behavior is primarily due to the design of the variational ansatz, rather than an intrinsic limitation of the PPS method itself.

In particular, we employ a class of parameterized circuits known as the Hamiltonian variational ansatz. The intuition behind this ansatz is as follows. To prepare a target state from a given initial state, where both are ground states of some Hamiltonians, one can imagine adiabatically tuning the Hamiltonian parameters, interpolating from those of the initial Hamiltonian to those of the target Hamiltonian. The Hamiltonian variational ansatz corresponds to a Trotterized version of this adiabatic evolution, with the Trotter steps promoted to variational parameters.

Heuristically, this approach performs well when the spectral gap remains open along the adiabatic path. However, its performance deteriorates, or requires a deeper circuit, when the gap shrinks or closes, as occurs near a critical point or in a gapless regime.

\subsection{Tilted-field quantum Ising model}\label{subsec: ising 1d gz}
Next, we consider the case $g_z \neq 0$. In this regime, the low-energy physics of the model is effectively described by the Ising field theory~\cite{delfinoSpinspin1995,fonsecaIsing2001a}, making the Hamiltonian a practical setting for studying scattering phenomena on quantum hardware.
In such applications, preparing the interacting ground state is a crucial prerequisite, whether for simulating scattering processes~\cite{milstedCollisions2022,jhaRealTime2024,farrellDigital2025,bennewitzSimulating2025} or exploring quantum quench dynamics~\cite{banulsStrong2011,kormosRealtime2017,linQuasiparticle2017}. Consequently, the preparation of the ground state via a quantum circuit plays a central role in enabling these quantum simulations.

\begin{figure}[!htbp]
    \centering
    {\includegraphics[width=0.235\textwidth]{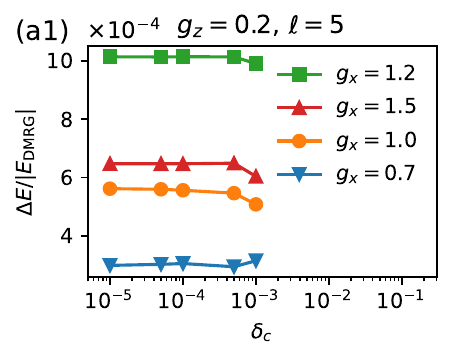}
    \includegraphics[width=0.235\textwidth]{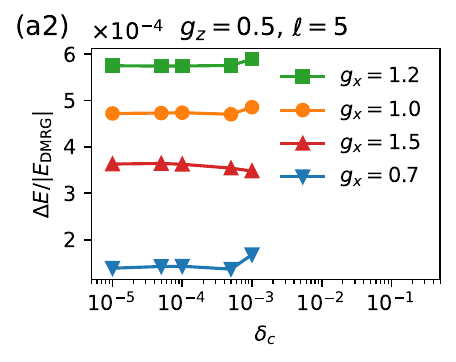}\\
    \includegraphics[width=0.235\textwidth]{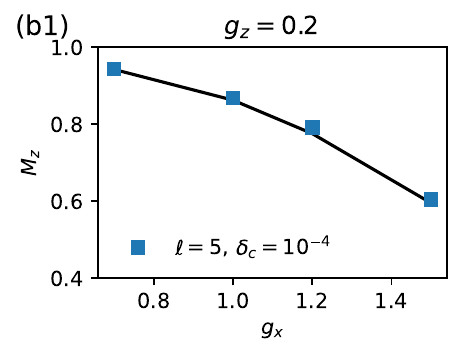}
    \includegraphics[width=0.235\textwidth]{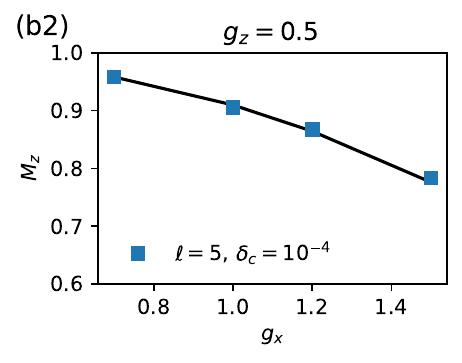}\\
    \includegraphics[width=0.235\textwidth]{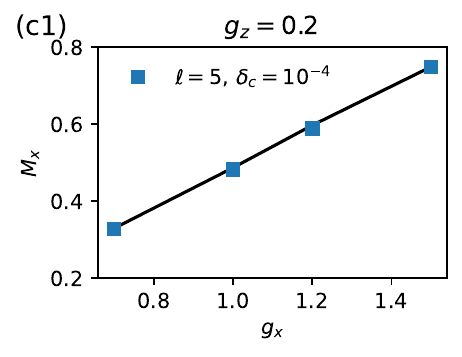}
    \includegraphics[width=0.235\textwidth]{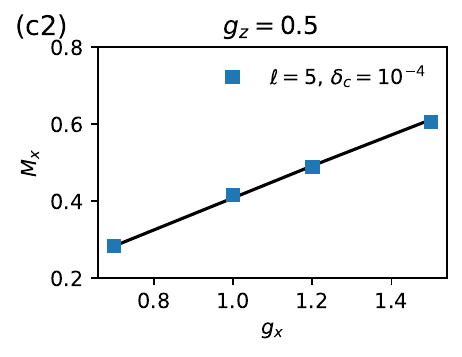}
    }
 
    \caption{\label{fig:1dising gz nonzero}
    For the 1D quantum Ising model with PBC and $N=100$, the relative energy error $\Delta E /|E_{\text{DMRG}}|$ for (a1) $g_z=0.2$ and (a2) $g_z=0.5$, where $\Delta E = \langle H \rangle - E_{\text{DMRG}}$ and circuit repetition $\ell =5$. (b1)(b2) $M_z$ and (c1)(c2) $M_x$ evaluated for the same parameters at $\delta_c = 10^{-4}$. All results show good agreement with the corresponding DMRG data.}

\end{figure}

Since the model with $g_z \neq 0$ is not exactly solvable, we use the ground-state energy obtained from DMRG $E_{\text{DMRG}}$ as the benchmark.
We define the relative error in this case as $\Delta E / |E_{\text{DMRG}}|$, where $\Delta E = \la \hat{H} \ra -E_{\text{DMRG}}$. 
In Figs.~\ref{fig:1dising gz nonzero} (a1) and (a2), we plot the relative error for $g_{z} = 0.2$ and $g_{z} = 0.5$, respectively, and for various $g_x$, using a repetition $\ell = 5$ with $N=100$.
Again, the optimization is carried out using PPS with $\delta_c = 10^{-3}$ typically.
Note the scale difference from Fig.~\ref{fig:1dsing_gz0}(a) --- the relative error remain less than $0.1\%$ here. 
Figures~\ref{fig:1dising gz nonzero}(b) and (c) show the expectation values of $M_z$ and $M_x$, respectively, which also exhibit strong agreement with DMRG results across the parameter range studied.

It is worth emphasizing that for gapped 1D Hamiltonians, such as Eq.~(\ref{eqn: 1d Ising}) with $g_z \neq 0$, DMRG is guaranteed to obtain the ground state with high precision using modest bond dimensions.
However, to study scattering or quench dynamics on quantum hardware, the ground state must be prepared on the device. 
While there exist methods to prepare matrix-product states on quantum hardware~\cite{schonSequential2005,malzPreparation2024}, the circuit depth and gate complexity can become substantial for modest bond dimensions, posing practical challenges.
Our parametrized circuit approach offers an alternative route for efficient ground-state preparation. As demonstrated in our results, even a shallow circuit with repetition $\ell = 5$ achieves remarkably small relative errors, highlighting the potential of this approach for quantum simulations.

\section{Quantum Ising model in 2D}\label{sec: ising 2D}

\begin{figure}[!tbp]
    \centering
    {\includegraphics[width=0.2\textwidth]{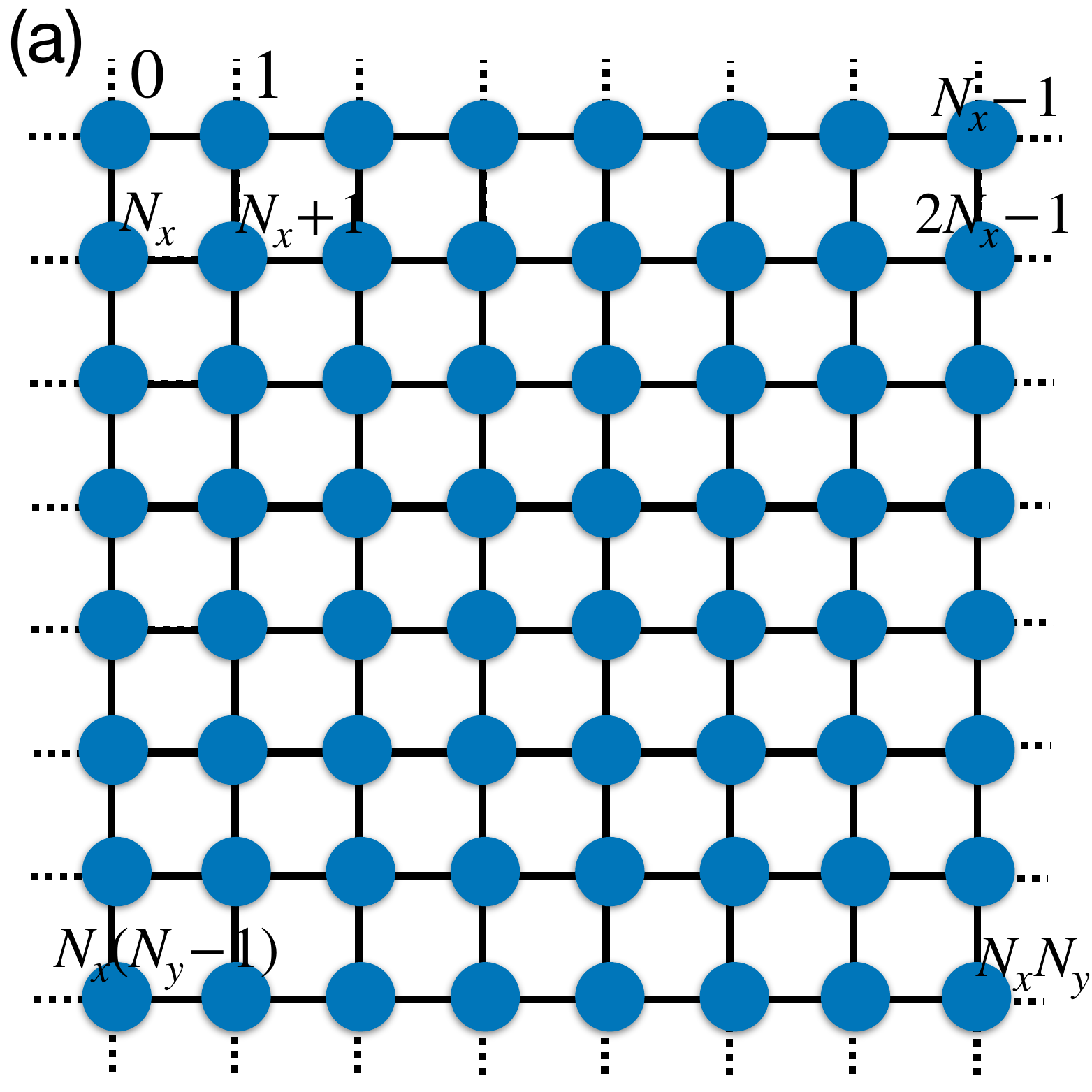}
    \includegraphics[width=0.265\textwidth]{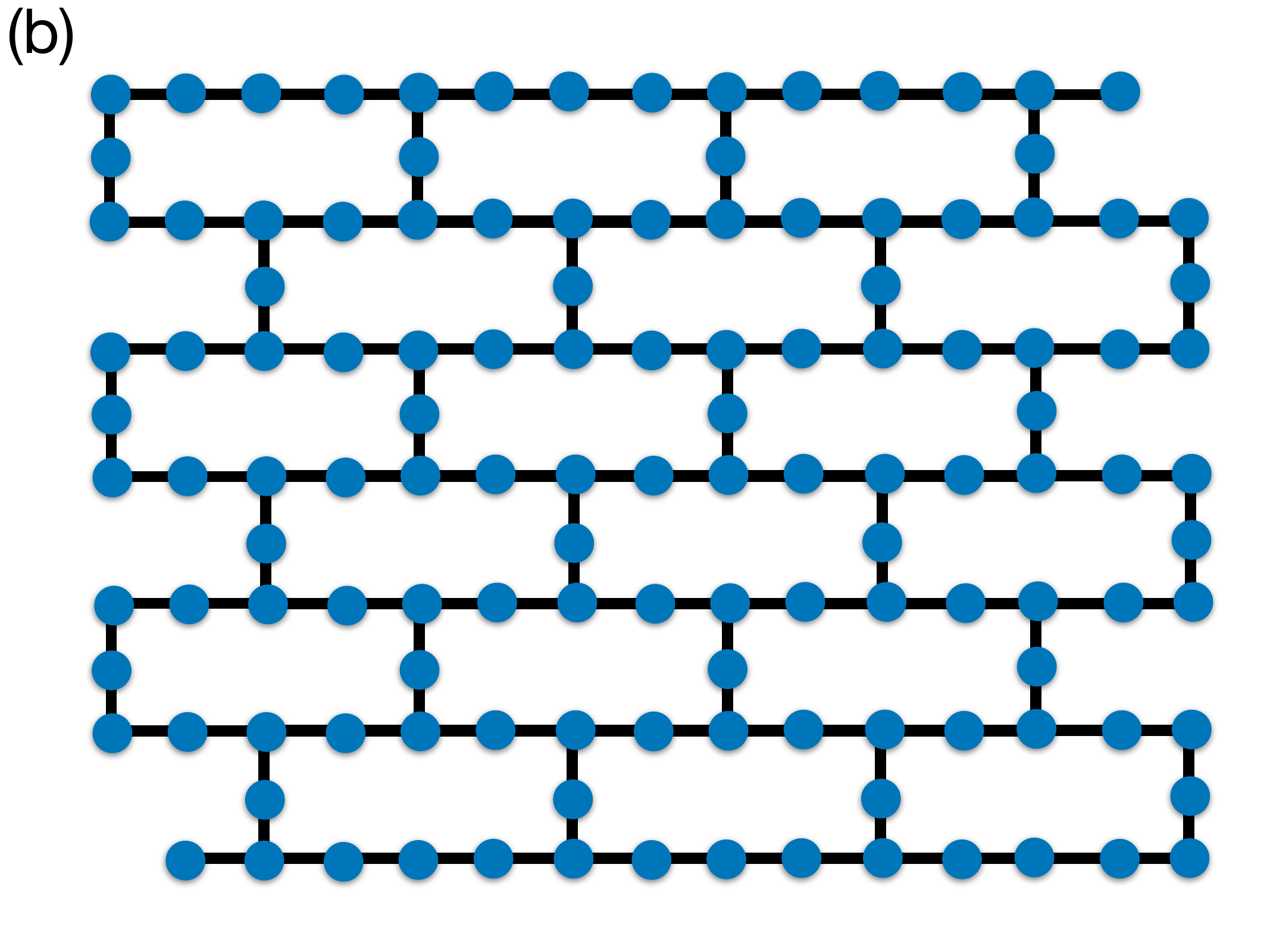}
    }
 
    \caption{\label{fig:2d_ising_lattices} (a) The square lattice with periodic boundary conditions. The lattice consists of a total of $N = N_x \times N_y$ qubits, where $N_x$ and $N_y$ represent the number of qubits in the $x$ and $y$ directions, respectively. The qubits are labeled according to the order shown in the figure.
    (b) The heavy-hex lattice with $N=127$ qubits.
    }
    \vspace{1em}
        {\includegraphics[width=0.48\textwidth]{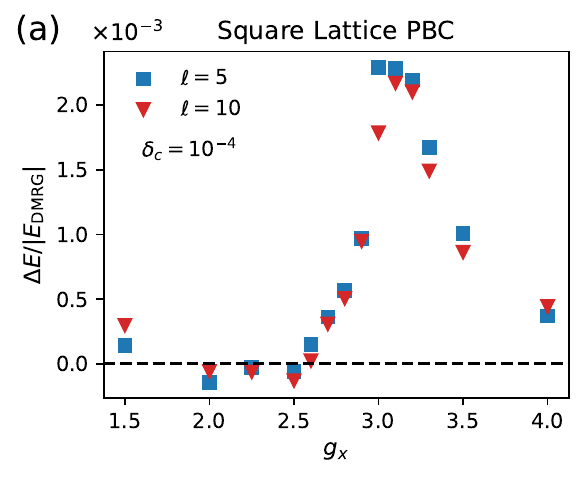}\\
    \includegraphics[width=0.235\textwidth]{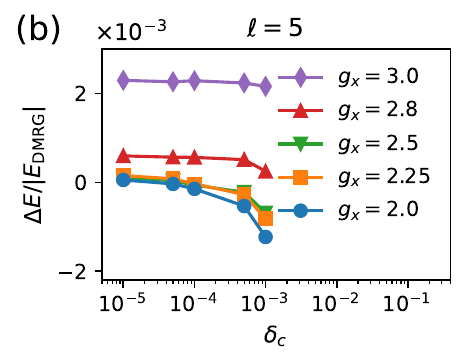}
    \includegraphics[width=0.235\textwidth]{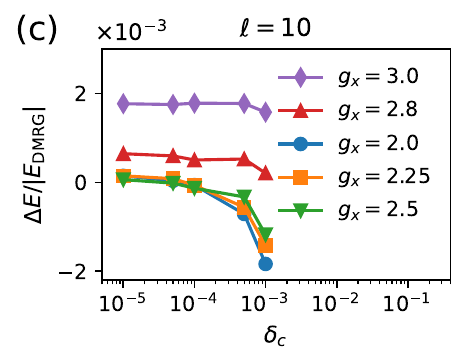}
    }
    \caption{\label{fig:2d_ising_rel_error_PBC} (a) The relative energy error $\Delta E / |E_{\text{DMRG}}|$ for the 2D quantum Ising model on the square lattice with PBC and $N_x=N_y=10$, where $\Delta E = \langle H \rangle - E_{\text{DMRG}}$. The variational energies $\langle H \rangle$ are re-evaluated at $\delta_c = 10^{-4}$, with circuit repetitions $\ell = 5$ and $\ell = 10$.
    (b)(c) Dependence of the relative energy error on the truncation threshold $\delta_c$ for circuit repetitions $\ell = 5$ and $\ell = 10$, respectively.}

\end{figure}

While DMRG is provably efficient for obtaining ground states of short-range interacting models in 1D, it is expected to face significant challenges in two or higher dimensions. It is therefore of particular interest to test and compare the performance of PPS parametrized circuits and DMRG on 2D models.
To this end, we study the quantum Ising model in 2D. Specifically, we consider a square lattice with PBC and a heavy-hex lattice with open boundary conditions (OBC), as illustrated in Figs.~\ref{fig:2d_ising_lattices}(a) and (b), respectively. Remarkably, as we will show later, the PPS parametrized circuit might outperform DMRG on the 2D quantum Ising model on a heavy-hex lattice in certain parameter regimes.

The Hamiltonian of 2D quantum Ising model is given as 
\begin{equation}\label{eqn: 2d Ising Ham}
H=-\sum_{\la i,j \ra}Z_iZ_{j}-g_x\sum_{j=1}^NX_j~,
\end{equation}
where $\la i,j \ra$ denotes the nearest-neighbor pairs of qubits, determined by the lattice geometries shown in Figs.~\ref{fig:2d_ising_lattices}(a) and (b).
The Hamiltonian has a $\mathbf{Z}_2$ symmetry generated by $\prod_{j=1}^N X_j$. As in the 1D case, we expect that the system will exhibit spontaneous symmetry breaking into a ferromagnetic phase when $g_x$ is small, while for large $g_x$ it enters the paramagnetic phase.

We use the following parameterized circuit to approximate the ground state, similar to the ansatz employed for the 1D quantum Ising model.
Define
\begin{equation}\label{eqn:2D_ising circuit_1}
    U(\alpha,\beta,\gamma)=u_{Z}(\alpha)u_{X}(\beta)u_{ZZ}(\gamma)~,
\end{equation}
where 
\begin{align}\label{eqn:2D_ising circuit_2}
    u_{Z}(\alpha) &= \prod_{j=1}^{N}\exp(-i\frac{\alpha}{2}Z_j) \notag \\
    u_{X}(\beta) &= \prod_{j=1}^{N}\exp(-i\frac{\beta}{2}X_j) \notag \\
    u_{ZZ}(\gamma) &= \prod_{\la i,j \ra}\exp(-i\frac{\gamma}{2}Z_iZ_{j})~.
\end{align}
The variational wavefunction of circuit repetition $\ell$ is again given by Eq.~(\ref{eqn: ising hva wavefunction}), and the cost function is given by Eq.~(\ref{eqn:cost function Ham}) with the corresponding Hamiltonian in Eq.~(\ref{eqn: 2d Ising Ham}).

We emphasize that while the properties of the 2D quantum Ising model can be calculated classically using quantum Monte Carlo algorithms~\cite{riegerApplication1999}, such methods do not yield a quantum circuit for approximately preparing its ground state. This is precisely the problem addressed by the parametrized circuit approach.

\subsection{Square lattice with PBC}

\begin{figure*}
    \centering
    {\includegraphics[width=0.48\textwidth]{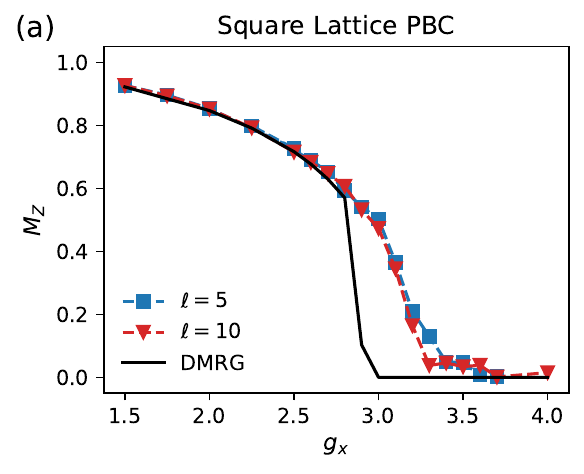}
    \includegraphics[width=0.48\textwidth]{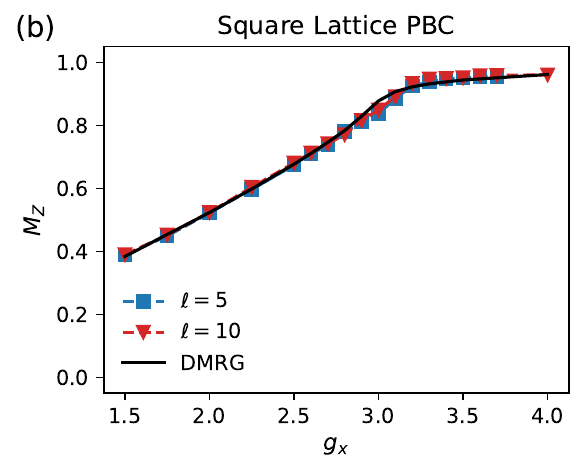}
    }
    \caption{\label{fig:2dising_pbc_Mz_Mx} For the 2D quantum Ising model on a square lattice with PBC and $N_x=N_y=10$, (a) $M_z$ and (b) $M_x$ for the variational wavefunction with circuit repetitions $\ell =5$ and $\ell =10$, evaluated at $\delta_c=10^{-4}$. The results show good agreement with the DMRG data away from the critical point $g_c \approx 3$. Near the critical point, we expect the results can be improved by increasing the circuit repetition.}

\end{figure*}

We first present results for the 2D quantum Ising model on a square lattice with PBC. We apply our method to a system of size $N_x = N_y = 10$, corresponding to a total of $N = 100$ qubits. Owing to the translational invariance of the variational wavefunction, the evaluation of the cost function can be simplified by applying PPS to the reduced Hamiltonian
$\tilde{H} = N \big( Z_0 Z_1 - Z_0 Z_{N_x} - g_x X_1 \big)$.
As a benchmark, we compare against results obtained using DMRG, implemented via snake-ordering (see Appendix~\ref{app: snake-ordering DMRG}), which maps the 2D lattice problem onto a 1D chain with long-range interactions.

Figure~\ref{fig:2d_ising_rel_error_PBC}(a) shows the relative error compared with DMRG energies obtained using maximum bond dimension $\chi=500$. This choice of bond dimension appears to yield energies converged to at least eight significant digits, with each sweep taking about five to ten minutes on a personal laptop. For PPS, most circuit parameters are optimized using a truncation threshold of $\delta_c = 10^{-3}$, with some trained at smaller $\delta_c$ for improved accuracy; all results are re-evaluated at $\delta_c = 10^{-4}$ for the plot. Interestingly, for $g_x$ in the range $2.0$–$2.5$, the variational wavefunction yields energies that are apparently lower than the DMRG results.

Figures~\ref{fig:2d_ising_rel_error_PBC}(b) and (c) further examine the dependence of the relative error on the truncation threshold $\delta_c$ for selected values of $g_x$. We find that the relative error becomes positive once $\delta_c$ is decreased sufficiently, confirming that the apparent energy undershoot is an artifact of truncation. In all cases, the relative energy errors benchmarked against DMRG remain remarkably small, below $0.25\%$.

In Figs.~\ref{fig:2dising_pbc_Mz_Mx} (a) and (b), we show the magnetizations $M_z$ and $M_x$ computed from the optimized variational wavefunction and compare them with the corresponding DMRG results. As in the 1D case, the variational wavefunction tends to overestimate $M_z$ near the critical point, which we conjecture could be mitigated by increasing the circuit repetition. In contrast, deep in the ferromagnetic phase ($g_x \ll g_c \approx 3$)~\cite{riegerApplication1999} and in the paramagnetic phase ($g_x \gg g_c \approx 3$), the variational results closely track the DMRG benchmarks.

\subsection{Heavy-hex lattice with OBC}
The second lattice geometry we study is the heavy-hex lattice with OBC, shown in Fig.~\ref{fig:2d_ising_lattices}(b), whose connectivity matches that of IBM quantum devices. We approximate the ground state of the quantum Ising model on this lattice using the parametrized circuit defined in Eqs.~(\ref{eqn:2D_ising circuit_1}) and (\ref{eqn:2D_ising circuit_2}), adapted to the heavy-hex connectivity, and benchmark the results against DMRG calculations with a maximum bond dimension of $\chi = 500$.

\begin{figure*}
    \centering
  \begin{minipage}[c]{0.65\columnwidth}
    \centering
    \includegraphics[width=0.65\linewidth]{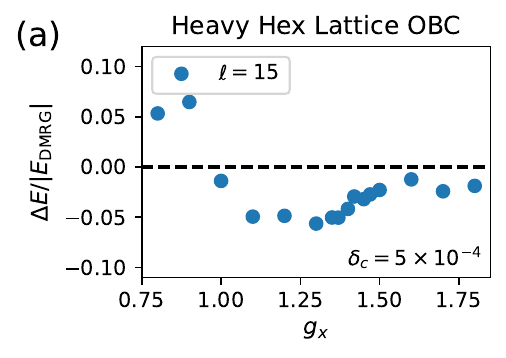}\\
    \includegraphics[width=0.65\linewidth]{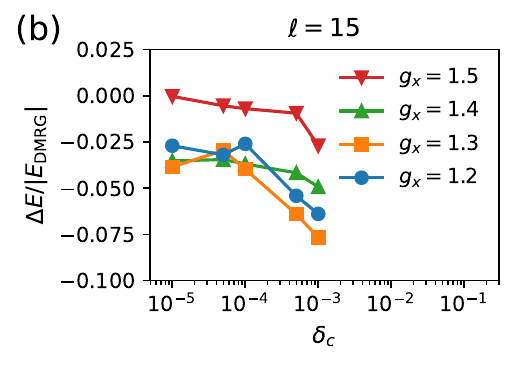}
  \end{minipage}
    \hspace{-3em}
\begin{minipage}[c]{0.75\columnwidth}
    \centering
    \includegraphics[width=1\linewidth]{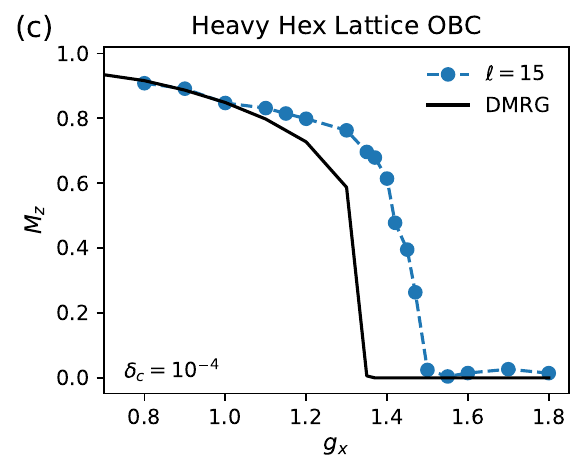}
  \end{minipage}
  \hspace{-0.5em}
  \begin{minipage}[c]{0.75\columnwidth}
    \centering
    \includegraphics[width=1\linewidth]{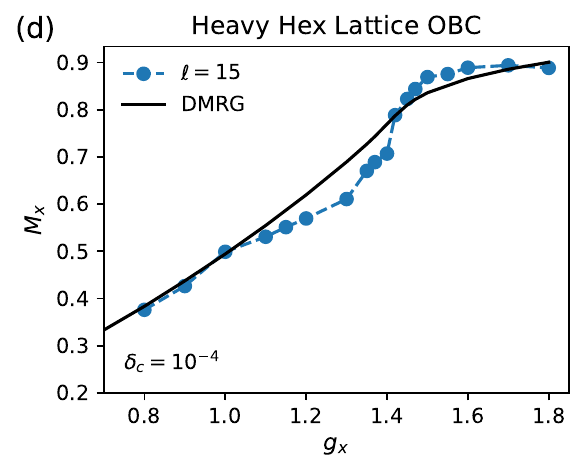}
  \end{minipage}

    \caption{\label{fig:2dising_obc} For the 2D quantum Ising model on the heavy-hex lattice with OBC and $N = 127$, (a) the relative energy error $\Delta E / |E_{\text{DMRG}}|$, where $\Delta E = \langle H \rangle - E_{\text{DMRG}}$, using the variational energies $\langle H \rangle$ re-evaluated at $\delta_c =5 \times 10^{-4}$ and circuit repetition $\ell = 15$.
    The PPS result achieves variational energy lower than the DMRG benchmark across a broad range of the transverse field strength $g_x$.
    (b) The dependence of the relative energy error on the truncation threshold $\delta_c$ for selected values of $g_x$. For some parameters, the variational energy remains below the DMRG energy even at $\delta_c=10^{-5}$. 
    (c)(d) Expectation values of the magnetization $M_z$ and $M_x$ compared to the DMRG results.}

\end{figure*}

In Fig.~\ref{fig:2dising_obc}(a), we plot the relative error compared with the DMRG result. The circuit parameters are mostly trained using $\delta_c = 10^{-3}$ but are re-evaluated with $\delta_c = 5\times 10^{-4}$ in Fig.~\ref{fig:2dising_obc}(a). Notably, across a broad range of parameters, the variational energies obtained from the PPS-trained parametrized circuit are lower than those from DMRG.
In Fig.~\ref{fig:2dising_obc}(b), we further examine the relative error as a function of $\delta_c$ for selected values of $g_x$. Remarkably, around $g_x \approx 1.3$, the PPS energy remains lower than the DMRG energy even at the smallest threshold, $\delta_c = 10^{-5}$.

However, we caution and remark the implication of this result.
First, the DMRG benchmarks reported here were obtained with modest computational effort on a personal laptop. The DMRG energy could likely be improved by adopting a different qubit-to-1D mapping, increasing the bond dimension, or performing additional sweeping iterations. Nevertheless, we believe the reported values are representative of what non-experts can reasonably achieve using the ITensor library~\cite{itensor} without extensive optimization of the DMRG algorithm. The PPS variational results were obtained with a comparable level of computational effort.

Second, as discussed earlier, truncation in the PPS algorithm can artificially lower the variational energy. The extent of this artifact can be assessed by examining its scaling with $\delta_c$. In the present case, even at a stricter threshold of $\delta_c = 10^{-5}$, the PPS variational energy remains below the DMRG result, although we cannot rule out the possibility that an even smaller $\delta_c$ would raise the PPS energy above the DMRG value.

Third, DMRG may not be the optimal classical method for this problem. Other classical approaches, such as projected entangled pair states or quantum Monte Carlo, could potentially achieve lower trial energies. Moreover, incorporating the symmetries of the problem can further enhance the performance of these classical methods; this was not explored in the present work.

Despite these caveats, it is striking that PPS variational results can yield lower variational energies than DMRG under these conditions. This observation suggests an intriguing potential application for quantum devices: validating whether such variational wavefunctions indeed achieve lower energies than the DMRG benchmark. The presented parametrized circuit, with repetition $\ell=15$ on the heavy-hex lattice with $N=127$ qubits, involves $2160$ two-qubit gates before transpilation to native gates. Definitive confirmation, however, would require quantum simulations under low-noise conditions or with extensive error mitigation, which lies beyond the present scope and resources of this work.

In Fig.~\ref{fig:2dising_obc}(c) and (d), we show the results for the magnetizations $M_z$ and $M_x$, respectively. We again observe that the variational wavefunction tends to yield higher $M_z$ values compared to the DMRG result near the critical point. Although we did not attempt to precisely locate the critical point here, Fig.~\ref{fig:2dising_obc}(c) suggests it lies close to $g_x \approx 1.4$.

We emphasize that, although the ground-state properties of the quantum Ising model on both lattice geometries can be studied classically using quantum Monte Carlo~\cite{riegerApplication1999} or tensor-network-based methods such as projected entangled pair states, these approaches do not produce explicit quantum circuits for approximately preparing the ground state. In contrast, our PPS parametrized circuit approach directly yields such circuits.

Moreover, our results demonstrate that PPS variational results can achieve high-quality approximations of ground states, underscoring their potential as a quantum-inspired classical numerical method for quantum many-body problems in higher dimensions. Unlike quantum Monte Carlo, this approach is applicable even to models where the sign problem poses a major obstacle. 
Its promise is further highlighted by our observation that PPS variational energies might, in some regimes, outperform DMRG, as shown for the heavy-hex lattice, though this should be interpreted with caution. 
Finally, in regimes where PPS encounter limitations, the PPS-trained circuit parameters can be directly transferred to quantum hardware for further refinement, thereby reducing both runtime and resource requirements on quantum devices.

\section{Kitaev honeycomb model}\label{sec: kitaev honeycomb}

\begin{figure}
    \centering
    \includegraphics[width=0.48\textwidth]{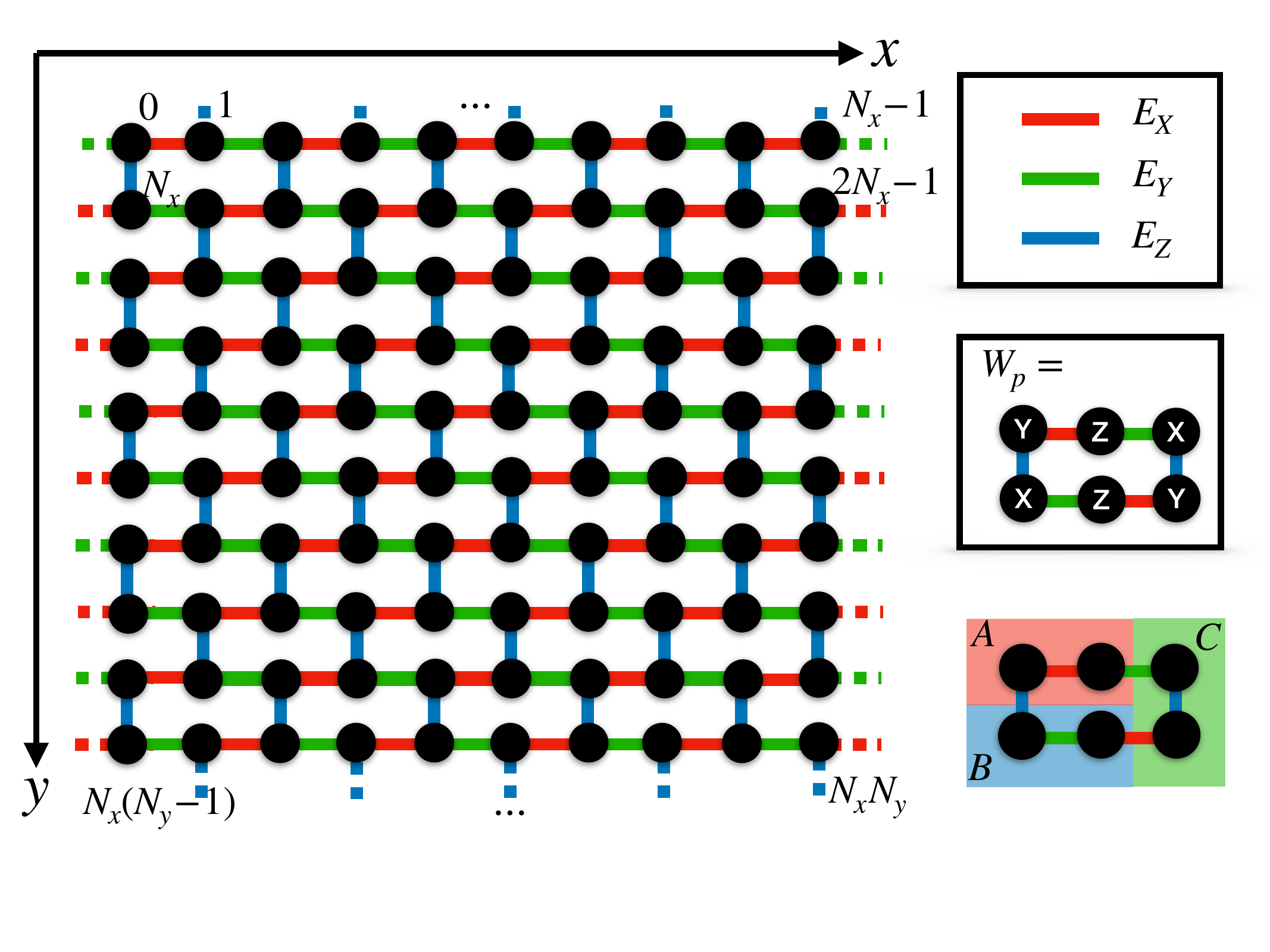}

 \caption{\label{fig:honeycomb_lattices} The honeycomb lattice can be mapped onto a square lattice with modified connectivity, as illustrated. It consists of $N = N_x \times N_y$ qubits, where $N_x$ and $N_y$ denote the number of qubits along the $x$- and $y$-directions, respectively. The colored bonds represent the $x$-bonds $E_X$ (red), $y$-bonds $E_Y$ (green), and $z$-bonds $E_Z$ (blue). Insets: The Kitaev honeycomb model has the conserved quantities $W_p$ shown in the figure. For the calculation of the topological entanglement entropy, we use the partition $A$, $B$, and $C$ shown as illustrated.
}

\end{figure}

The last model to which we apply our method is the Kitaev honeycomb model~\cite{Kitaev2005}, for which several works have proposed quantum circuits to prepare its ground state~\cite{xiaoDetermining2021,bespalovaQuantum2021,jahinFermionic2022,liBenchmarking2023,parkDigital2025,aliRobust2025a}. For convenience, we map the honeycomb lattice onto a square lattice with modified connectivity with PBC, as illustrated in Fig.\ref{fig:honeycomb_lattices}. The bonds are color coded by type: red for $x$-type bonds ($E_X$), green for $y$-type bonds ($E_Y$), and blue for $z$-type bonds ($E_Z$). The Hamiltonian is
\begin{equation}\label{eqn: kitaev_honeycomb Ham}
H = -J_x\sum_{\langle i,j\rangle \in E_X}  X_i X_j - J_y\sum_{\langle i,j\rangle \in E_Y}  Y_i Y_j - J_z\sum_{\langle i,j\rangle \in E_Z}  Z_i Z_j,
\end{equation}
where $J_x$, $J_y$, and $J_z$ are the coupling strengths along the $x$-, $y$-, and $z$-type bonds, respectively, and the sums run over nearest-neighbor qubit pairs $\langle i, j \rangle$ of the corresponding bond type.

This model is exactly solvable, making it an ideal testbed for benchmarking ground-state energy estimates. 
Notably, the Kitaev honeycomb model has conserved quantities $W_P$, often referred to as fluxes, on each hexagonal plaquette as illustrated in the inset of Fig.~\ref{fig:honeycomb_lattices}.
In the thermodynamic limit, the ground states lies in the ``flux-free" sector, where $W_P = +1$ for all plaquettes.
The exact ground state can be obtained using the Majorana parton construction, as described in Ref.~\cite{Kitaev2005}. (See also Ref.~\cite{chenExact2008} for an alternative approach based on the Jordan-Wigner transformation.)

The Kitaev honeycomb model exhibits two distinct quantum phases, known as the A phase and the B phase, depending on the relative strengths of the bond couplings $J_x$, $J_y$, and $J_z$.
The system is in the gapped A phase when one coupling dominates, such as when $|J_z| > |J_x| + |J_y|$. In this regime, the model realizes a gapped quantum spin liquid with Abelian anyonic excitations.
In contrast, the gapless B phase arises when the couplings satisfy $|J_z| \leq |J_x| + |J_y|$, $|J_y| \leq |J_x| + |J_z|$, and $|J_x| \leq |J_y| + |J_z|$.
The gaplessness in this phase originates from the massless Majorana fermion excitations, while the flux excitations ($W_p = -1$) are still gapped. 
In our study, we fix $J_z = 1$ and vary $J_x = J_y = J$ from $0$ to $1$. Under this parametrization, the system lies in the gapped A phase when $J < 0.5$, and in the gapless B phase for $J \geq 0.5$.

\begin{figure*}
    \centering

  \begin{minipage}[c]{0.82\columnwidth}
    \centering
    \includegraphics[width=1\linewidth]{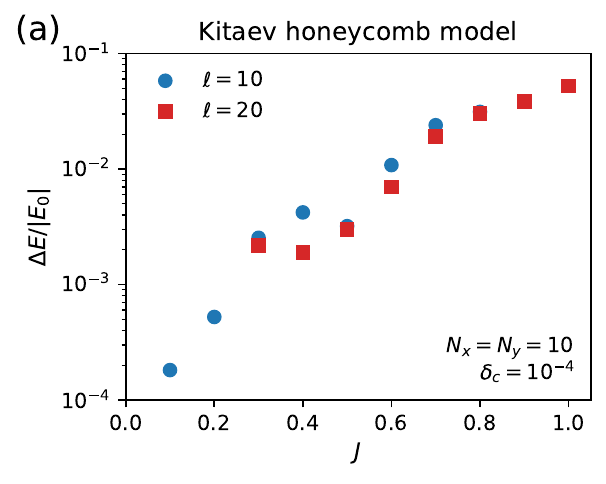}
  \end{minipage}
  \hspace{-3.5em}
  \begin{minipage}[c]{0.65\columnwidth}
    \centering
    \includegraphics[width=0.6\linewidth]{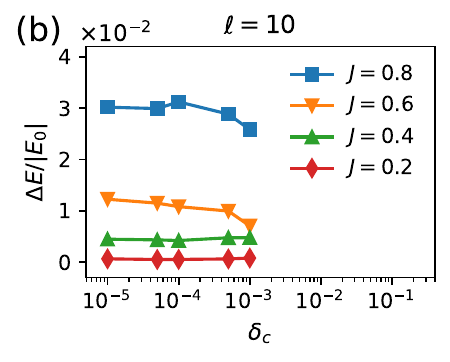}\\
    \includegraphics[width=0.6\linewidth]{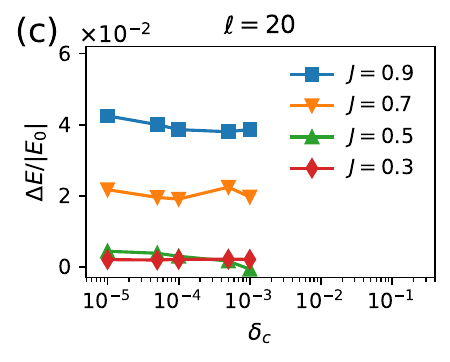}
  \end{minipage}
  \hspace{-3.5em}
  \begin{minipage}[c]{0.82\columnwidth}
    \centering
    \includegraphics[width=1\linewidth]{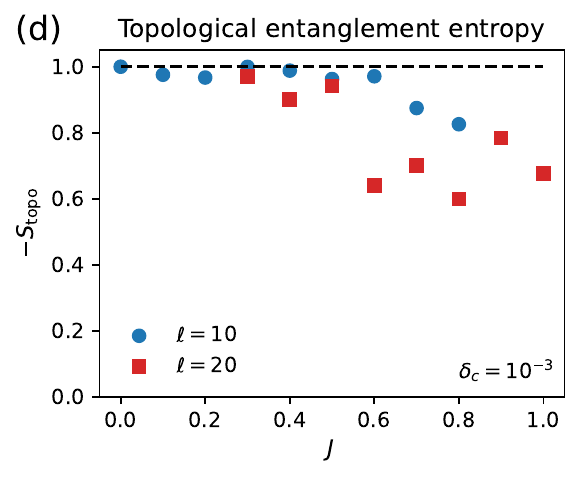}
  \end{minipage}

    \caption{\label{fig:kitaev honeycomb} For the Kitaev honeycomb model with $N_x=N_y=10$ and $J_z=1$. (a) The relative energy error $\Delta E/|E_0|$, where $\Delta E = \la H \ra - E_0$, for parameters $J_x=J_y=J$ and circuit repetitions $\ell =10$ and $\ell =20$, evaluated at $\delta_c=10^{-4}$.
    Note the logarithmic scale in the $y$-axis. The relative energy error remains lower than $0.5\%$ for $J \leq 0.5$.
    (b)(c) The dependence of the relative energy error on $\delta_c$ for ansatz depths $\ell = 10$ and $\ell = 20$, respectively. 
    (d) The (negative) topological entanglement entropy $-S_{\text{topo}}$ computed using the partition $A =\{0,1\}$, $B=\{ N_x,N_x+1\}$ and $C=\{ 2,N_x+2\}$, evaluated at $\delta_c = 10^{-3}$. The dashed line at $-S_{\text{topo}} =1$ marks the theoretical value for the Kitaev A phase in the limit where the regions $A$, $B$ and $C$ are large.}

\end{figure*}

To construct the quantum circuit for preparing the ground state of the Kitaev honeycomb model, we employ the following parametrized circuit:
\begin{equation}
U(\alpha,\beta,\gamma)=u_{ZZ}(\alpha)u_{YY}(\beta)u_{XX}(\gamma)~
\end{equation}
where
\begin{align}
    u_{ZZ}(\alpha) &= \prod_{\la i,j \ra \in E_Z}\exp(-i\frac{\alpha}{2}Z_iZ_j)~, \notag \\
    u_{YY}(\beta) &= \prod_{\la i,j \ra \in E_Y}\exp(-i\frac{\beta}{2}Y_iY_j)~, \notag \\
    u_{XX}(\gamma) &= \prod_{\la i,j \ra \in E_X}\exp(-i\frac{\gamma}{2}X_iX_j)~.
\end{align}
The resulting variational wavefunction with a circuit repetition  $\ell$ is
\begin{equation}\label{eqn:kitaev-hva-wavefunction}
|\psi(\boldsymbol{\alpha},\boldsymbol{\beta},\boldsymbol{\gamma})\rangle
= U(\alpha_\ell,\beta_\ell,\gamma_\ell)\cdots U(\alpha_1,\beta_1,\gamma_1)V_{\text{f.f.}}|0\rangle^{\otimes N}~,
\end{equation}
where $V_{\text{f.f.}}$ is a circuit that prepares the “flux-free” state, satisfying $W_p=+1$ for all plaquettes and $Z_iZ_j=+1$ for all $i,j$ on the $z$-type bonds $E_Z$. This particular flux-free state also corresponds to a ground state of the Kitaev honeycomb model at $J_x=J_y=0$ and $J_z=1$.
While the full description of $V_{\text{f.f.}}$ is provided in Appendix~\ref{app: fluxfree}, we note here that it consists entirely of Clifford gates, ensuring that it does not increase the number of Pauli terms in PPS. The circuit has a depth approximately equal to the linear dimension of the system, scaling as $N_x + N_y$.

Note that while the parameterized circuit $U(\alpha,\beta,\gamma)$ is translationally invariant, the initial state preparation circuit $V_{\text{f.f.}}$ and the PPS truncation break the translation symmetry. 
Nevertheless, we find that the resulting variational wavefunction remains nearly translationally invariant in practice.
To leverage this approximate symmetry, we optimize the circuit parameters using the proxy Hamiltonian $\tilde{H} = \frac{1}{2}N_xN_y(-J_x X_0 X_1 - J_y Y_{N_x}Y_{N_x+1}-J_z Z_{2}Z_{N_x+2})$ and $\delta_c = 10^{-3}$.
We find that using the proxy Hamiltonian improves the quality of the optimization, as the initial operator in PPS now involves only $3$ Pauli-strings, compared to $\sim 1.5N$ number of Pauli-strings if using the full Hamiltonian.

In Fig.~\ref{fig:kitaev honeycomb}(a), we show the relative error benchmarked against the exact ground state energy $E_0$, for a system size $N_x=N_y =10$, or total $N =100$ qubits. 
Note that the relative errors are re-evaluated using $\delta_c = 10^{-4}$ and computed with respect to the full Hamiltonian Eq.~(\ref{eqn: kitaev_honeycomb Ham}) rather than the proxy.
We find that for $J < 0.5$, corresponding to the gapped A phase, the relative errors remain below $0.5\%$ for the largest repetition $\ell$ considered. 
We observe that the optimization becomes  more challenging when $J \geq 0.5$, where the system enters the gapless B phase.
Figures~\ref{fig:kitaev honeycomb} (b) and (c) show the relative error as a function of $\delta_c$ for selected $J$.

As mentioned previously, the Kitaev A phase is a gapped topological state with Abelian anyonic excitations. 
A key fingerprint of such topological order is the topological entanglement entropy~\cite{kitaevTopological2006,levinDetecting2006}.
We consider a subregion of the system partitioned into three parts A, B, and C as shown in the inset of Fig.~\ref{fig:honeycomb_lattices}. 
The topological entanglement entropy is defined as 
\begin{equation}
    S_{\text{topo}}=S_A+S_B+S_C-S_{AB}-S_{AC}-S_{BC}+S_{ABC}~,
\end{equation}
where $S_{R} := -\tr[\rho_{R}\log_2 \rho_{R}]$ is the von Nuemann entanglement entropy of the reduced density matrix $\rho_R := \tr_{R^c}[|\psi\ra \la \psi|]$, and $R^c$ denotes the complement of region R. 
We use abbreviations such as $AB := A \cup B$, and so on.
For sufficiently large regions A, B, and C,  the topological entanglement entropy in the Kitaev A phase takes the value $S_{\text{topo}}=-1$.

We compute the topological entanglement entropy for the optimize variational wavefunction. Specifically, we take the regions $A =\{1,2\}$, $B=\{N_x,N_{x+1}\}$, and $C=\{2,N_x+1\}$, reconstructing the reduced density matrix $\rho_{ABC}$ via state tomography using PPS.
To do this, we express $\rho_{ABC}$ in a Pauli-string expansion
\begin{equation}
    \rho_{ABC} = \frac{1}{2^{|ABC|}}\sum_{\hat{P}} c_{\hat{P}} \hat{P}~,
\end{equation}
where $\hat{P} \in \{I,X,Y,Z\}^{\otimes |ABC|}$ runs over the Pauli strings in the region $ABC$, and each coefficient is given by $c_{\hat{P}} = \la \psi(\boldsymbol{\theta}) |\hat{P}|\psi(\boldsymbol{\theta})\ra$, evaluated using PPS.
Since the number of Pauli strings grows exponentially with subsystem size, we limit to a subsystem with $|ABC| = 6$ qubits, a size for which PPS tomography remains computationally feasible using $\delta_c= 10^{-3}$.

Fig.~\ref{fig:kitaev honeycomb} (d) shows the result of the topological entanglement entropy computed from the variational wavefunctions at various values of $J$, for repetitions $\ell=10$ and $\ell=20$. 
The data point at $J = 0$, $\ell = 10$ corresponds to the state $V_{\text{f.f.}}|0\ra^{\otimes N}$, which has $S_{\text{topo}} = -1$.
Remarkably, for $J \leq 0.5$, $S_{\text{topo}}$ remains close to the theoretical value of $-1$, despite the relatively small sizes of regions A, B, and C. On the other hand, $S_{\text{topo}}$ starts to deviate from $-1$ for $J > 0.5$.   
These results highlight the quality of our variational wavefunction in approximating the ground states of Eq.~(\ref{eqn: kitaev_honeycomb Ham}).

\section{Simulation on quantum hardware}\label{sec: quantum validation}

In the previous sections, we obtained the circuit parameters using PPS.
Since these parameters were obtained via a noiseless, classical approximation of the cost function, it is natural to ask how well the resulting states perform when the parameterized circuit is executed on a noisy near-term quantum device. 
To this end, we consider the Kitaev honeycomb model and the variational wavefunction described in Sec.~\ref{sec: kitaev honeycomb} on a system of size $(N_x, N_y) = (8, 6)$, corresponding to a total of $N = 48$ qubits. We focus on two representative coupling values, $J = 0.3$ and $J = 0.6$, with repetition $\ell=5$. After obtaining the optimized parameters using PPS at $\delta_c = 10^{-3}$ with respect to the proxy Hamiltonian, we prepare the PPS-trained state on the Quantinuum H2-2 quantum computer~\cite{quantinuum2025}.

Note that the initial part of the trial state $V_{\text{f.f}}|0\ra^{\otimes N}$ is a stabilizer state, which in principle could be prepared using measurement and feedforward, resulting in a constant-depth circuit~\cite{iqbalTopological2024}. 
In our quantum simulations, however, we implement this part of the circuit explicitly using the quantum gates described in Appendix~\ref{app: fluxfree}, which has a circuit depth scaling as $N_x + N_y$. 
For circuit repetition $\ell = 5$, the total number of two-qubit gates is approximately $500$ prior to transpilation into the native gate set of the Quantinuum H2-2 quantum computer. It would be an interesting direction for future work to investigate whether preparing $V_{\text{f.f}}|0\ra^{\otimes N}$ via measurement and feedforward could yield a higher-quality approximate ground state on the quantum device.

\begin{figure*}
    \centering

 \begin{minipage}[c]{0.68\columnwidth}
    \centering
    \includegraphics[width=\linewidth]{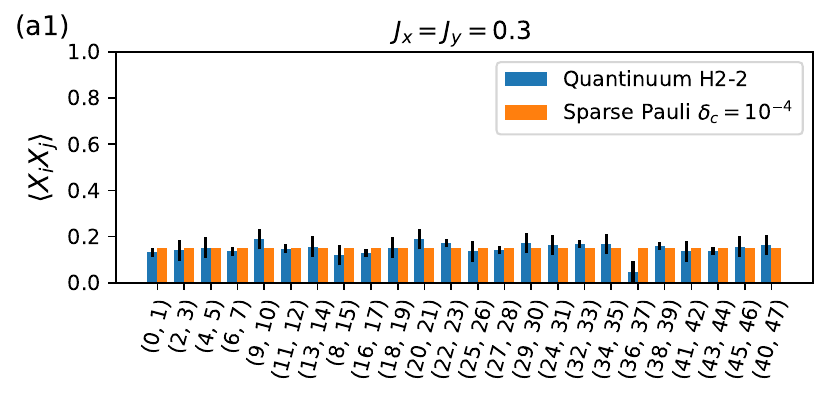}\\
    \includegraphics[width=\linewidth]{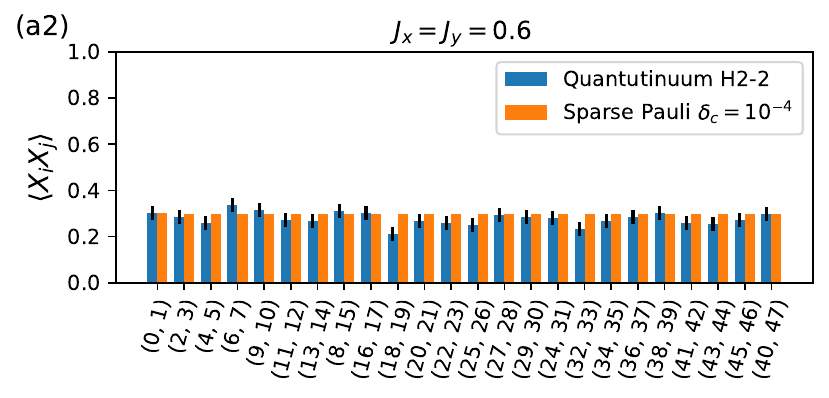}
  \end{minipage}
  \hspace{-.5em}
  \begin{minipage}[c]{0.68\columnwidth}
    \centering
    \includegraphics[width=\linewidth]{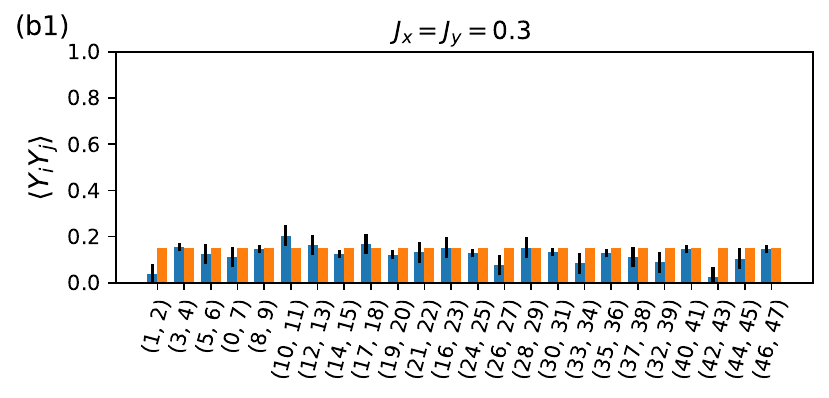}\\
    \includegraphics[width=\linewidth]{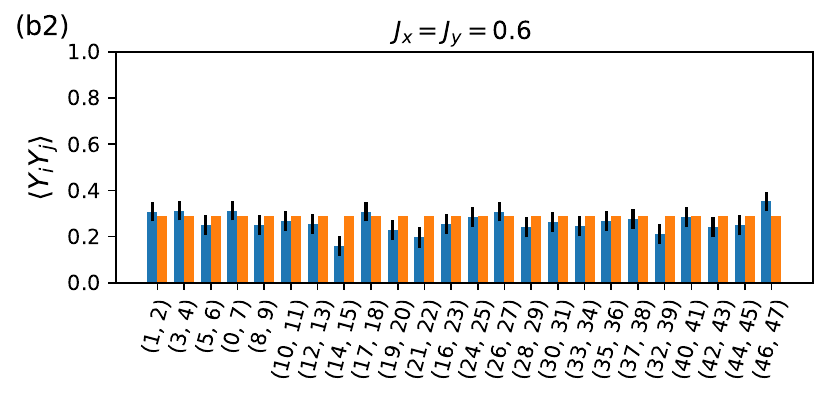}
  \end{minipage}
  \hspace{-.5em}
 \begin{minipage}[c]{0.68\columnwidth}
    \centering
    \includegraphics[width=\linewidth]{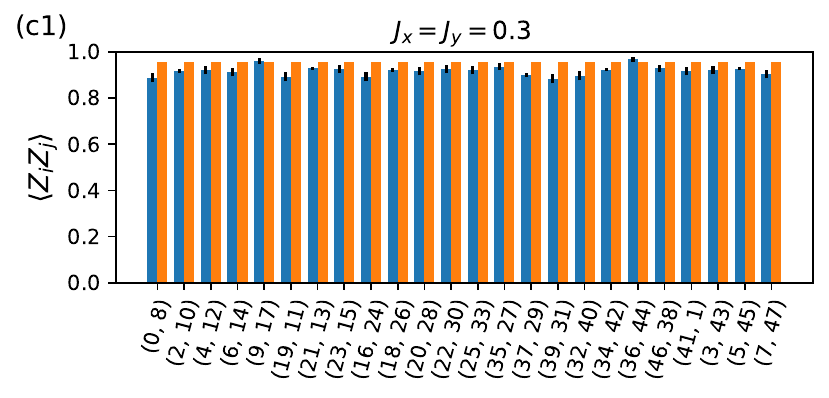}\\
    \includegraphics[width=\linewidth]{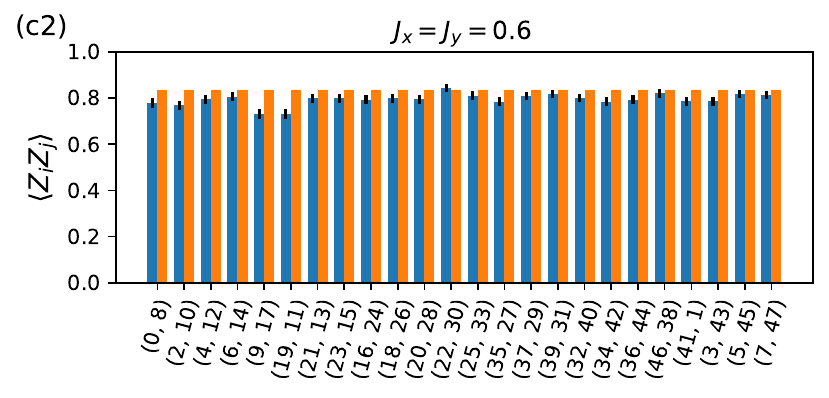}
  \end{minipage}

    \caption{\label{fig:quantinuum}
    For the Kitaev honeycomb model with $(N_x,N_y)=(8,6)$ and $J_z =1$, the expectation values of (a1) $\la X_iX_j \ra$ on $x$-bonds, (b1) $\la Y_iY_j \ra$ on $y$-bonds, and (c1) $\la Z_iZ_j \ra$ on $z$-bonds, for $J_x=J_y=0.3$, obtained from both the Quantinuum H2-2 quantum computer and PPS method. 
    (a2)(b2)(c2) Similar to (a1)(b1)(c1) but for the couplings $J_x=J_y=0.6$.}

\end{figure*}

\subsection{Energy}\label{subsec: Quantinuum energy}
We first examine the energy of the approximate ground state prepared on the Quantinuum H2-2 quantum computer.
In Fig.~\ref{fig:quantinuum}, we show the measured values of $\la X_iX_j \ra$, $\la Y_iY_j \ra$, and $\la Z_iZ_j \ra$ on the corresponding bonds from the Quantinuum machine, alongside the results from the PPS method. 
The Quantinuum data are collected by executing the parametrized circuit and performing measurements in the $x$, $y$, and $z$ bases. The resulting bitstrings from each basis are then post-processed to compute the expectation values $\langle X_i X_j \rangle$, $\langle Y_i Y_j \rangle$, and $\langle Z_i Z_j \rangle$. For $J = 0.3$, most bonds are measured with 500 shots, while selected bonds are measured with 3000 shots; for $J = 0.6$, measurements are taken with 1000 shots for $x$- and $z$-type bonds, and 500 shots for $y$-type bonds. The error bars are estimated from the standard error 
$s /\sqrt{N_{\text{shot}}}$, where $N_{\text{shot}}$ is the number of shots and $s = \sqrt{\sum_{j=1}^{N_{\text{shot}}}(x_j - \bar{x})^2/(N_{\text{shot}}-1)}$ is the sample standard deviation. 

As noted earlier, the variational wavefunction is approximately translationally invariant. This is evident in the PPS results, where the values  of $\la X_iX_j \ra$ are nearly identical across all the $x$-bonds, and similarly for the $y$-bonds and $z$-bonds.
The Quantinuum measurements show overall good agreement with the PPS predictions, apart from a few bonds that show noticeable deviations.
From the measured $\la X_iX_j \ra$, $\la Y_iY_j \ra$, and $\la Z_iZ_j \ra$, we compute 
\begin{align}
    \la H \ra =& -J_x\sum_{\la i,j\ra\in E_X}\la X_iX_j\ra -J_y\sum_{\la i,j\ra\in E_Y}\la Y_iY_j\ra  \notag \\
    &-J_z\sum_{\la i,j\ra\in E_Z}\la Z_iZ_j\ra~, \notag
\end{align}
which gives the energy estimation on the Quantinuum machine
\begin{equation}
    E_Q= -23.9717 \pm 0.0539 \notag
\end{equation}
for $J=0.3$ and 
\begin{equation}
    E_Q= -26.8396 \pm 0.0893 \notag
\end{equation}
for $J=0.6$. The quoted uncertainties are estimated via standard error propagation under the assumption that all bond measurements are statistically independent. 
Although a more accurate estimate should be obtained from $\sqrt{(\la H^2 \ra - \la H \ra^2)/(N_{\text{shot}}-1)}$, where the covariance among the bonds are included, we expect the independent-bond assumption to be a reasonable approximation in this context.

For comparison, the exact ground-state energies are
\begin{equation}
    E_0 = -25.0873 \notag
\end{equation} for $J=0.3$ and 
\begin{equation}
    E_0 = -28.2847 \notag
\end{equation}
for $J=0.6$, yielding relative energy errors $\Delta E/|E_0| = 4.45\%$ for $J=0.3$ and $\Delta E/|E_0| = 5.11\%$ for $J=0.6$. 
These results are obtained without any error mitigation. Notably, even in the presence of hardware noise, our parameterized circuit can prepare states roughly $5\%$ in relative energy error for a system of $N = 48$ qubits.

The PPS energies evaluated at $\delta_c =10^{-5}$ with respect to the full Hamiltonian are
\begin{equation}
    E_{\text{PPS}} = -25.0906 \notag
\end{equation} for $J=0.3$ and 
\begin{equation}
    E_{\text{PPS}} = -28.3715 \notag
\end{equation}
for $J=0.6$.
We again observe the artifact that truncation can lead to a trial energy slightly below the true ground-state energy. 
A summary of energies obtained from the different methods is provided in Table~\ref{tab:energy}.

\begin{table}
    \centering
    \begin{tabular}{c||c|c|c}
         $J$ & $E_0$ & PPS & Quantinuum H2-2\\
         \hline
         $0.3$ & $ -25.0873$ & $ -25.0906$ & $ -23.9717 \pm 0.0539$ \\
         $0.6$ & $ -28.2847$ & $ -28.3715$ & $ -26.8396 \pm 0.0893$ \\
    \end{tabular}
 
    \caption{Comparison of the ground-state energies $E_0$ of the Kitaev honeycomb model at $J = 0.3$ and $J = 0.6$ on a system of size $(N_x, N_y) = (8, 6)$, with the corresponding results from PPS and Quantinuum H2-2 quantum computers.}
    \label{tab:energy}
\end{table}

\subsection{Braiding statistics}\label{sec:pps braiding}
In addition to benchmarking the energy, we investigate whether the quantum states prepared on the device exhibit the topological properties of the Kitaev honeycomb model by examining the braiding statistics of the anyonic excitations. 
In this model, the expected anyonic excitations are $e$, $m$, and $\psi$. 
For the lattice shown in Fig.~\ref{fig:honeycomb_lattices}, we label plaquette excitations with $W_P = -1$ in the odd rows as $e$ anyons and those in the even rows as $m$ anyons. The $\psi$ anyons correspond to the underlying Majorana fermions in the Majorana parton construction~\cite{Kitaev2005}, where a pair of $\psi$ excitations can be created by applying a string of operators consisting of $XX$, $YY$, or $ZZ$ on the corresponding $E_X$, $E_Y$, or $E_Z$ bonds.

\begin{figure*}[t]
 \begin{minipage}[c]{0.68\columnwidth}
    \includegraphics[width=\linewidth]{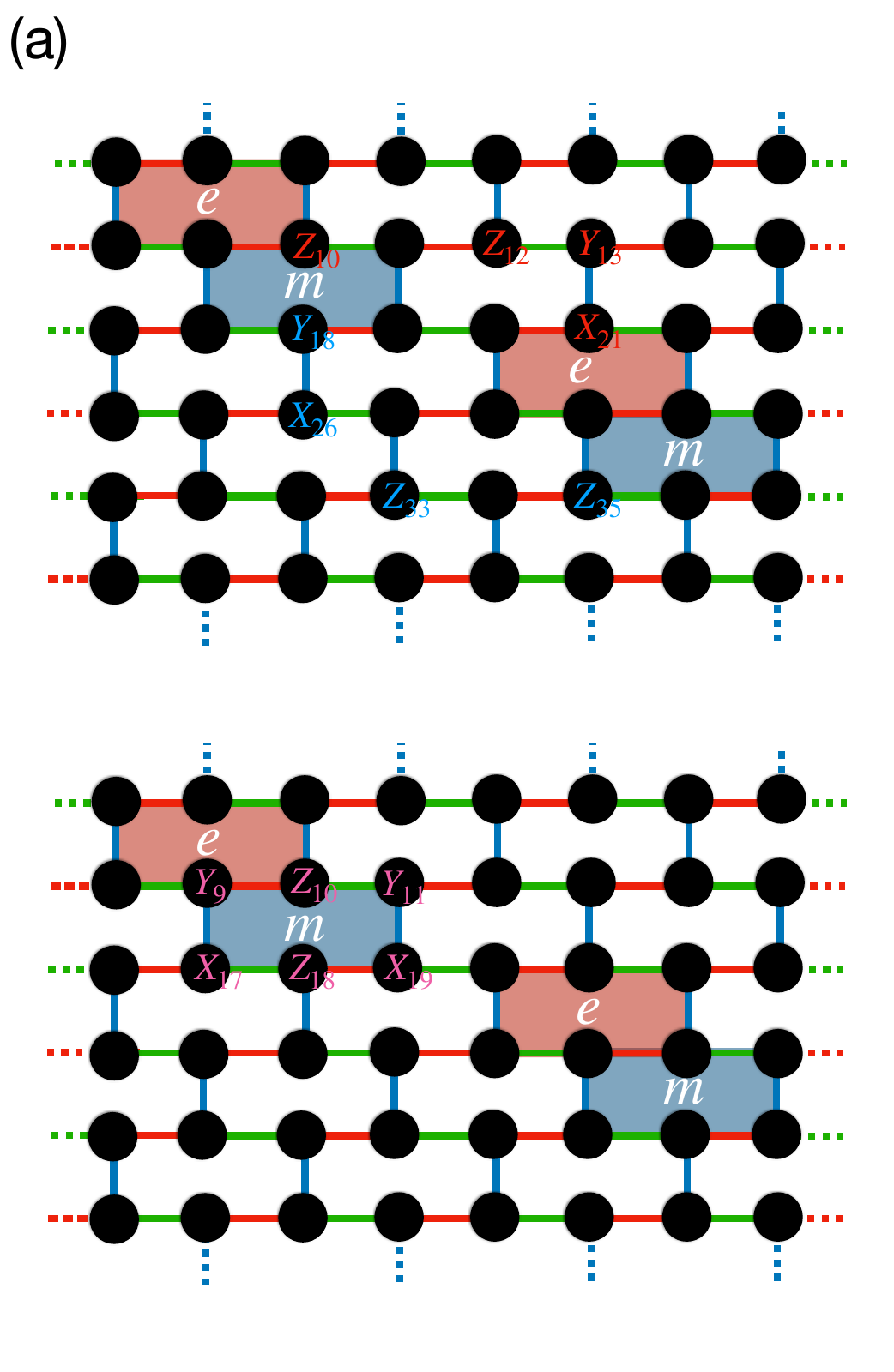}
  \end{minipage}
  \hspace{-.5em}
  \begin{minipage}[c]{0.68\columnwidth}
    \centering
    \includegraphics[width=\linewidth]{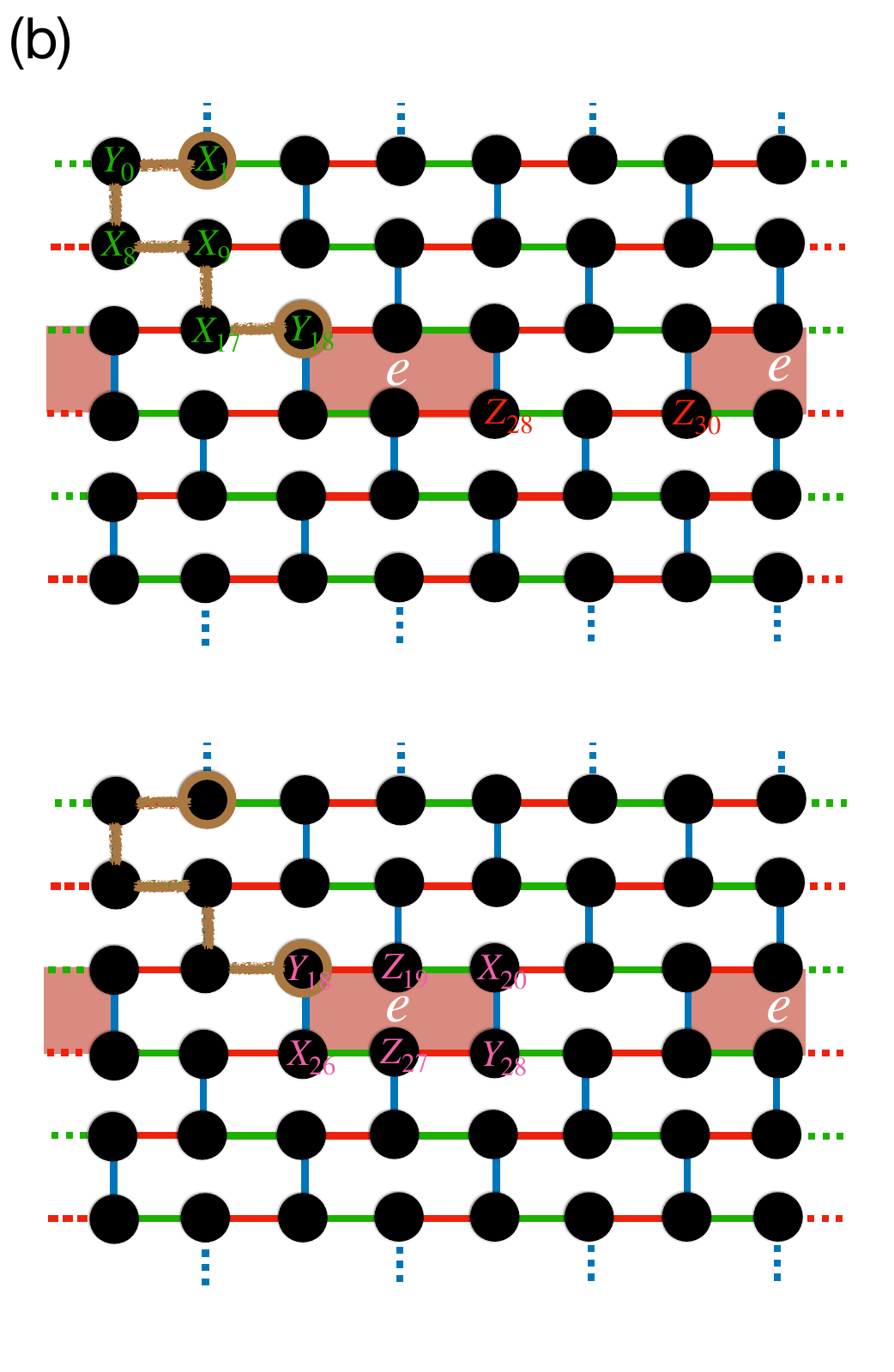}
  \end{minipage}
  \hspace{-.5em}
 \begin{minipage}[c]{0.68\columnwidth}
    \centering
    \includegraphics[width=\linewidth]{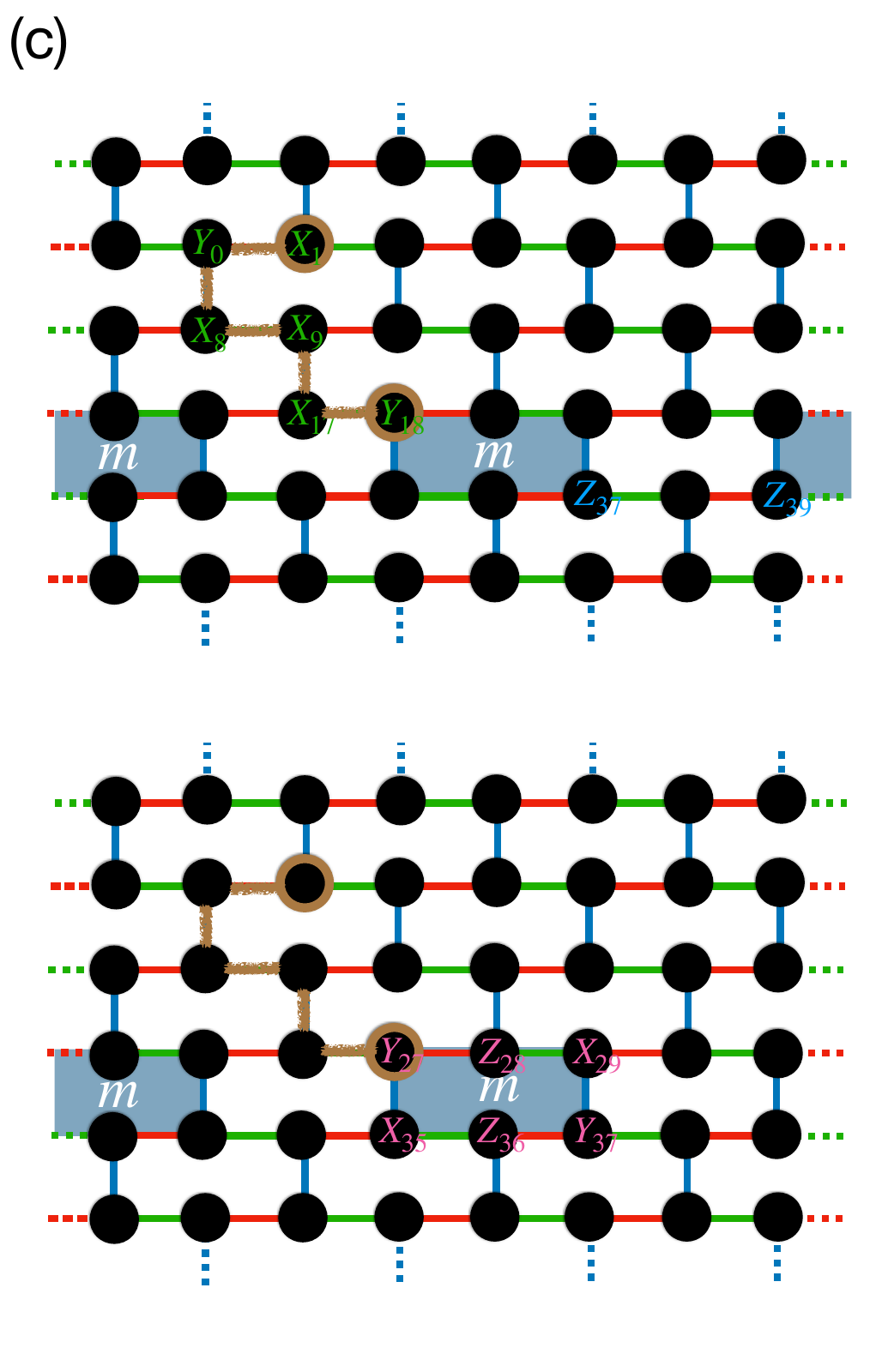}
  \end{minipage}

    \caption{\label{fig:braiding}
    For the Kitaev honeycomb model with $(N_x,N_y)=(8,6)$, the anyon creation operations (top panels) and the braiding operations (bottom panels) used to extract the braiding statistics.
    (a) $e$-$m$ braiding: the $e$ anyons are created by the gates shown in red, while the $m$ anyons are created by the gates shown in blue. The braiding operation, implemented with the gates in purple, can be interpreted as moving the top-left $e$ anyon around the nearby $m$ anyon.
    (b) $e$-$\psi$ braiding: the $\psi$ anyons are created by the string of gates shown in brown, with the pair of $\psi$ excitations residing at the ends of the string, highlighted by the brown circles. The braiding operation, again implemented with the purple gates, moves the central $\psi$ anyon around the nearby $e$ anyon.
    (c) $m$-$\psi$ braiding: analogous to the $e$-$\psi$ case, where the purple gates implement a braiding operation that moves the central $\psi$ anyon around the nearby $m$ anyon.
    }
\end{figure*}

The Abelian nature of these anyons implies that braiding one anyon around another results in an additional phase imprinted on the quantum state. To detect this phase, we adopt the interferometric protocol introduced in Ref.~\cite{satzingerRealizing2021}. 
At a high level, the protocol begins by preparing two pairs of anyons of different types on top of the ground state, yielding a state we denote as $|\Phi\rangle$. 
An ancilla qubit is then initialized in the superposition state $H|0\rangle_a = |+\rangle_a = \tfrac{1}{\sqrt{2}}(|0\rangle + |1\rangle)_a$. 
A controlled-braiding operation is applied to $|\Phi\rangle$, with the ancilla serving as the control qubit. 
More explicitly, if the braiding unitary is $U_b$, the controlled braiding maps the joint state as
$|\Phi\rangle \otimes |+\rangle_a \;\mapsto\; \tfrac{1}{\sqrt{2}}\big(|\Phi\rangle \otimes |0\rangle_a + U_b|\Phi\rangle \otimes |1\rangle_a\big)$.
The expectation values of the ancilla in the $X$ and $Y$ bases then yield the real and imaginary parts of the braiding phase, respectively
\begin{align}
\langle X_a \rangle = \re\big(\langle \Phi|U_b|\Phi\rangle\big), \quad
\langle Y_a \rangle = \im\big(\langle \Phi|U_b|\Phi\rangle\big).
\end{align}

\begin{table}[!htbp]
        \begin{minipage}{1\linewidth}
        \raggedright
        (a) $e$-$m$ braiding
        \resizebox{\columnwidth}{!}{
        \begin{tabular}{|c ||  c | c|}
        \hline
        $J$  & PPS & Quantinuum H2-2 \\
        \hline 
        $0.3$  & $-0.9501$ & $(-0.8400 \pm 0.0243) + i(-0.0360 \pm 0.0447)$\\
        $0.6$  & $-0.8141$ & $(-0.7160 \pm 0.0313) + i(0.0440 \pm 0.0447)$ \\
        \hline 
       \end{tabular}
       }
       \end{minipage}
        \vspace*{0.3cm}

        \begin{minipage}{1\linewidth}
        \raggedright
        (b) $e$-$\psi$ braiding
        \resizebox{\columnwidth}{!}{
        \begin{tabular}{|c ||  c | c|}
        \hline 
        $J$  & PPS & Quantinuum H2-2\\
        \hline 
        $0.3$  & $-1.0000$ & $(-0.9160  \pm 0.0180) + i(-0.0960 \pm 0.0446)$\\
        $0.6$  & $-1.0000$ & $(-0.9240 \pm  0.0171) + i(0.0760 \pm 0.0446)$ \\
        \hline 
       \end{tabular}
       }
       \end{minipage}
       \vspace*{0.3cm}
       
       \begin{minipage}{1\linewidth}
       \raggedright
        (c) $m$-$\psi$ braiding
        \resizebox{\columnwidth}{!}{
        \begin{tabular}{|c ||  c | c|}
        \hline
        $J$  & PPS & Quantinuum H2-2 \\
        \hline 
        $0.3$  & $-1.0000$ & $(-0.9360 \pm  0.0158) + i(-0.0440 \pm 0.0447)$\\
        $0.6$  & $-1.0000$ & $(-0.9080 \pm 0.0188) + i(0.0600  \pm 0.0447)$ \\
        \hline 
       \end{tabular}
       }
       \end{minipage}

     \caption{The braiding statistics obtained from the PPS and Quantinuum H2-2 quantum computer, for which all phases are expected to be $-1$ theoretically.  The quantum simulation results have remarkable agreement with the theoretical value.}
     \label{tab:braiding}
\end{table}

We use the creation and braiding operations defined on the fixed-point state $V_{\text{f.f.}}|0\ra^{\otimes N}$ (corresponding to $J=0$) and apply them to the approximate ground states at $J=0.3$ and $J=0.6$ for a system of size ($N_x,N_y) = (8,6)$.
For the $e$–$m$ braiding, we apply the gates shown in the top panel of Fig.~\ref{fig:braiding} (a). A pair of $e$ anyons are excited by the red gates $Z_{10}Z_{12}Y_{13}X_{21}$, while a pair of $m$ anyons are excited by the blue gates $Y_{18}X_{26}Z_{33}Z_{35}$. These operations can be understood by examining their effect on the $W_P$ operators --- $Z$ operations excite two neighboring anyons or move an anyon to a neighboring plaquette in the $x$-direction, while the combined $YX$ operation moves an anyon across rows.
The braiding operation, depicted in the bottom panel of Fig.~\ref{fig:braiding} (a), consists of the purple gates, which effectively move the $e$ anyon at the top-left corner around the nearby $m$ anyon. 
The expected braiding phase is $\alpha_{em} = -1$.

We simulate the interferometry protocol both using PPS and the Quantinuum H2-2 quantum computer. For the fixed-point ($J=0$) state, the PPS result yields the expected $\alpha_{em} = -1$. For the variational states, PPS gives $\alpha_{em} = -0.9501$ at $J=0.3$ and $\alpha_{em} = -0.8141$ at $J=0.6$. 
The deviation from the theoretical value of $-1$ likely arises from the finite correlation length of the state, as well as the fact that the state is only an approximate ground state rather than the exact one.
On the quantum device, using $500$ shots per observable for $\langle X_a\rangle$ and $\langle Y_a \rangle$, we obtain
\begin{equation}
    \alpha_{em} = (-0.8400 \pm 0.0243) + i(-0.0360 \pm 0.0447) \notag
\end{equation}
for $J=0.3,$ and
\begin{equation}
    \alpha_{em} = (-0.7160 \pm 0.0313) + i(0.0440 \pm 0.0447) \notag
\end{equation}
for $J=0.6$, both without error mitigation. 
A summary of these results is given in Table~\ref{tab:braiding}(a).

For the $e$–$\psi$ braiding, the anyon preparation operation is shown in the top panel of Fig.~\ref{fig:braiding}(b), and the braiding operation in the bottom panel. To move a $\psi$ anyon across a bond, we apply $XX$, $YY$, or $ZZ$ gate depending on whether the bond belongs to $E_X$, $E_Y$, or $E_Z$, respectively. Thus, to move the $\psi$ anyon in the middle around the nearby $e$ anyon in a counterclockwise path, we apply
$(X_{18}X_{19})(Y_{19}Y_{20})(Z_{20}Z_{28})(X_{27}X_{28})(Y_{26}Y_{27})(Z_{18}Z_{26})
= Y_{18}X_{26}Z_{27}Y_{28}X_{20}Z_{19}$,
which is precisely the plaquette operator $W_p$.
The presence of an $e$ anyon on that plaquette makes $W_p=-1$.

Indeed, the PPS results for $J = 0,\, 0.3,\, 0.6$ all yield $\alpha_{e\psi} = -1$. On the Quantinuum device, we obtain
\begin{equation}
    \alpha_{e\psi} = (-0.9160 \pm 0.0180) + i(-0.0960 \pm 0.0446)~\notag
\end{equation}
for $J=0.3$, and
\begin{equation}
\alpha_{e\psi} = (-0.9240 \pm 0.0171) + i(0.0760 \pm 0.0446)    \notag
\end{equation}
for $J=0.6$, where the real and imaginary parts are each estimated using $500$ shots without error mitigation.
These results are summarized in Table~\ref{tab:braiding} (b).

Lastly, the $m$–$\psi$ braiding follows a protocol very similar to the $e$–$\psi$ case, with the anyon creation operation shown in the top panel of Fig.~\ref{fig:braiding}(c) and the braiding operation in the bottom panel. As before, the $\psi$ braiding effectively acquires a phase corresponding to $W_p = -1$, and the PPS results for $J = 0,\, 0.3,\, 0.6$ all yield $\alpha_{m\psi} = -1$.
On the quantum device, we obtain
\begin{equation}
    \alpha_{m\psi} = (-0.9360 \pm 0.0158) + i(0.0440 \pm 0.0447) \notag
\end{equation} for $J=0.3$ and 
\begin{equation}
    \alpha_{m\psi} = (-0.9080 \pm 0.0188) + i(0.0600 \pm 0.0447) \notag
\end{equation} 
for $J=0.6$, where the real and imaginary components are each estimated using $500$ shots without error mitigation.
These results are summarized in Table~\ref{tab:braiding} (c).

We remark that demonstrations of anyon braiding on quantum devices have so far primarily focused on fixed-point models~\cite{satzingerRealizing2021,niuDemonstrating2024,iqbalTopological2024,iqbalNonAbelian2024a,minevRealizing2025}. In contrast, our results demonstrate braiding statistics on ground states beyond the fixed point, showing that the quantum states prepared with our variational circuit on the device exhibit the expected topological properties. A summary of the braiding phases is provided in Table~\ref{tab:braiding}.

\section{Discussion}~\label{sec: discussion}
In this work, we propose using Pauli Path simulation for variational quantum algorithms in the task of ground-state preparation at the utility scale, i.e., for system sizes at which generic quantum circuits are beyond the reach of exact state-vector simulation. To test this approach, we apply it to the quantum Ising model in one dimension, in two dimensions on both a square lattice with periodic boundary conditions and a heavy-hex lattice with open boundary conditions, as well as to the Kitaev honeycomb model, all at system sizes of one hundred qubits or more. Across all test cases, we benchmark Pauli Path simulation results against either exact solutions or density-matrix renormalization group results, finding remarkable agreement as quantified by the relative energy error and certain observables. For the 2D quantum Ising model on a heavy-hex lattice, the Pauli Path-simulated circuit may even outperform density-matrix renormalization group in certain parameter regimes, albeit with caveats. Finally, we validate and benchmark our approach on Quantinuum H2-2 quantum computer~\cite{quantinuum2025} for the Kitaev honeycomb model. For a system size of $N = 48$ qubits, the prepared ground states achieve a relative energy error of roughly $5\%$ without any error mitigation. Moreover, the extraction of braiding statistics confirms that the prepared quantum states exhibit the expected topological properties.

Our Pauli Path-simulated variational quantum algorithm offers several advantages and potentials in both quantum computing application and classical algorithm. First, it provides a systematic approach for classically constructing parametrized circuits to approximately prepare the ground state of a system --- a prerequisite for executing tasks such as quench dynamics, scattering simulations on a quantum machine, or probing dynamical responses.  While certain models have ground states that are exactly solvable (e.g. the one-dimensional quantum Ising model and the Kitaev honeycomb model) or can be analyzed using classical numerical methods such as the density-matrix renormalization group (e.g. short-ranged interacting models in 1D) or quantum Monte Carlo methods (e.g. the two-dimensional quantum Ising model), preparing these states on a quantum machine can be challenging. 
The preparation procedure is not always straightforward, and even when a method exists --- such as when the classical description of the wavefunction is a matrix-product state~\cite{schonSequential2005,malzPreparation2024} --- state preparation can be resource-intensive. 
Our Pauli path-simulated variational quantum algorithm addresses these challenges by finding the quantum circuit through classical computation, which can then be executed on a quantum machine, as demonstrated in Sec.~\ref{sec: quantum validation}.

Second, our Pauli Path–simulated variational quantum algorithm can serve as a novel quantum-inspired classical numerical method for condensed matter physics, materials science, and even quantum chemistry. This potential has already been hinted at by its successful application to certain quantum machine learning problems~\cite{bermejoQuantum2024}. As demonstrated in this work, our method can approximate the ground states of various models with very small relative error, particularly when the models possess a finite energy gap. For the two-dimensional quantum Ising model on a heavy-hex lattice, our Pauli Path–based variational energy may even outperform results obtained from the density-matrix renormalization group, using comparable computational effort for both methods, though with important caveats. 

However, we caution that, due to the truncation scheme, the variational energy calculated from PPS may fall below the true ground-state energy, which could be problematic when no other reliable classical benchmark is available. In such cases, examining the scaling of the trial energy as a function of the truncation threshold can provide some confidence in the reliability of the result. Moreover, even if the PPS-computed trial energy becomes formally unphysical, the optimized parameters still define a valid trial wavefunction, which often performs well in practice. Evaluating the optimized circuit using other classical simulation methods, such as tensor-network approaches, or even executing it on a quantum device can serve as an independent validation of the variational energy. 

While the properties of quantum Ising models in any dimension can be analyzed using quantum Monte Carlo, our method extends to models where quantum Monte Carlo encounters difficulties, most notably those afflicted by the sign problem. Moreover, whereas the performance of density-matrix renormalization group is constrained by the amount of entanglement in the system, the Pauli Path method is instead limited by the system’s ``magic," which quantifies its degree of non-Clifford-ness. 
Systems whose ground states are both highly entangled and sign-problematic, yet exhibit low magic, may therefore represent a regime where our Pauli Path–simulated variational quantum algorithm outperforms existing classical approaches.

Moreover, the Pauli Path–simulated variational quantum algorithm is inherently quantum-hardware-friendly. 
In parameter regimes where the Pauli Path-simulated parametrized circuit alone becomes challenging, its results can be seamlessly transferred to a quantum machine to initialize and execute the variational quantum algorithm. 
In our Pauli Path–simulated optimizations, we observe that using circuit parameters obtained at nearby parameter points as a warm start significantly improves optimization efficiency. 
This strategy can, in principle, be applied directly on quantum hardware: circuit parameters obtained in regimes where the Pauli Path simulation performs well can serve as a warm start for the quantum-executed variational quantum algorithm, potentially reducing quantum runtime and enhancing the quality of the optimization results.

In the future, we expect the Pauli Path–simulated variational quantum algorithm to be further improved and applied to a wider range of problems.
One promising avenue is ansatz design, which is a crucial component of the variational quantum algorithm framework.
Our method provides a computationally inexpensive way to explore different ansätze on a much larger system size without requiring access to actual quantum hardware. 
In this work, we adopted a parameterized circuit based on the Hamiltonian variational ansatz, which is particularly well-suited for condensed matter problems. 
A natural extension would be to apply our method to adaptive ansatz constructions~\cite{grimsleyAdaptive2019}, which are often more appropriate for quantum chemistry  and quantum machine learning applications.

Another important direction is applying and testing our method on fermionic problems --- an essential step toward tackling realistic challenges in quantum materials and quantum chemistry. In such cases, one can employ various fermion-to-qubit encodings to map the fermionic Hamiltonian into a qubit Hamiltonian. 
Mappings such as the Jordan–Wigner transformation can introduce long-range or multi-qubit interactions, which pose difficulties for classical numerical methods and for near-term quantum hardware. 
These complications, however, are largely benign for the Pauli Path simulation.

Improving the parameter update algorithm is another technical challenge worth to pursue. In this work, we employed SPSA combined with ADAM. While SPSA has the advantage of requiring only two evaluations of the cost function per optimization step, its stochastic nature limits the final accuracy. 
This limitation can be mitigated by lowering the learning rate and by averaging the gradient estimates over multiple evaluations. Although we did not systematically test the optimization outcomes across many different random seeds, our limited test cases suggest that different randomized initial parameters lead to consistent final results up to several significant digits.

Exploring alternative optimization algorithms is therefore warranted. 
For instance, implementing automatic differentiation could reduce the number of cost function evaluations needed to compute gradients while maintaining high accuracy. 
Another promising option is the quantum natural gradient --- a second-order derivative based optimization method that has been shown to have advantage over first-order derivative based methods such as gradient descent or ADAM~\cite{wierichsAvoiding2020}. Incorporating the quantum natural gradient (or an approximate version of it) into our Pauli Path–simulated variational algorithm would be an interesting direction for future work.

Beyond algorithmic improvements, our method has the potential to raise the bar for claims of practical quantum advantage in the domain of variational quantum algorithms. As demonstrated in this work, Pauli Path–simulated results can achieve very low relative energy errors across a variety of quantum many-body Hamiltonians. To establish a genuine quantum advantage, a near-term device must therefore demonstrate that, despite the presence of noise, it can prepare a variational ground state with energy superior to what can be achieved classically. At the same time, our results highlight that systems with gapless energy spectra pose greater challenges for Pauli Path–simulated parametrized circuits. Determining whether quantum-executed variational algorithms can offer an advantage in such regimes is a worthwhile endeavor.

Finally, even when ground states can be approximated accurately by classical methods, their dynamical responses and real-time quantum dynamics are often far more difficult to simulate classically. This suggests a promising hybrid use case: preparing ground states via the PPS variational method, and then leveraging quantum hardware to simulate the subsequent dynamics. Such scenarios could provide a realistic pathway to demonstrating practical quantum advantage.

\begin{acknowledgments}
We thank the entire BlueQubit team for contributions to the BlueQubit SDK and, in particular, to the Pauli Path Simulators used for the classical simulations in this work. We are grateful to the Quantinuum team -- particularly its Startup Partner Program --- for sustained support and for facilitating access to the System Model H2 trapped-ion processor. We also thank Siddharth Hariprakash and Zuhair Mullath for valuable discussions and feedback. 
C.-J.~Lin acknowledges the support from the National Science Foundation (QLCI grant OMA-2120757).
\end{acknowledgments}

\onecolumngrid

\appendix
\section{Parameter update algorithm}\label{app: SPSA and ADAM}
In this appendix, we describe the optimization algorithm used for updating the variational parameters. Our approach combines a modified Simultaneous Perturbation Stochastic Approximation (SPSA) with the Adaptive Moment Estimation (ADAM) algorithm.

Denote the variational parameters by  $\boldsymbol{\theta} = (\theta_1,\cdots, \theta_p)$ and the cost function by $C(\boldsymbol{\theta})$.  In the variational optimization framework, the most basic parameter update strategy is gradient descent, which updates parameters using the gradient of the cost function 
\begin{equation}
    \boldsymbol{\theta}_{i+1} = \boldsymbol{\theta}_{i} -\eta \nabla C(\boldsymbol{\theta}_{i})~,
\end{equation}
where $\boldsymbol{\theta}_{i}$ represents the parameters at the $i$-th iteration, and $\eta$ is the learning rate hyperparameter, which we tune according to the problem, with a typical value $\eta = 0.001$.
However, computing the full gradient $\nabla C(\boldsymbol{\theta})$ can become prohibitively expensive when the number of parameters $p$ is large. For example, using the finite difference method to approximate the gradient requires evaluating each component as $[C(\boldsymbol{\theta} + \Delta_{\ell}) - C(\boldsymbol{\theta} - \Delta_{\ell})]/(2\Delta_{\ell})$, where $\Delta_{\ell}$ is a small perturbation in the $\ell$-th component of $\boldsymbol{\theta}$.
This would require $2p$ evaluations of the cost function per iteration, which is computationally intensive for high-dimensional parameter spaces.

SPSA alleviates this challenge by approximating the gradient using only two evaluations of the cost function, regardless of the number of parameters.
\begin{equation}\label{appeqn: SPSA gradient}
    \nabla C(\boldsymbol{\theta}) \approx \frac{C(\boldsymbol{\theta}+\boldsymbol{\Delta})-C(\boldsymbol{\theta}-\boldsymbol{\Delta})}{2\Delta}\boldsymbol{\Delta} :=\mathbf{G}(\boldsymbol{\theta})~,
\end{equation}
where $\boldsymbol{\Delta}=(\Delta_1, \cdots, \Delta_p)$ is a randomly chosen perturbation vector in parameter space, and $\Delta=\|\boldsymbol{\Delta}\|_2 := \sqrt{\sum_{\ell=1}^{p}\Delta_\ell^2}$ is its Euclidean norm.
In practice, we default the hyperparameter $\Delta = 0.005$, which can be tuned depending on the problem.  
To generate $\boldsymbol{\Delta}$, we first draw a random vector $\boldsymbol{\chi} = (\chi_1, \dots, \chi_p)$ with each component $\chi_\ell$ sampled uniformly from the interval $[-1, 1]$. 
We then normalize this vector and scale it to the desired norm $\boldsymbol{\Delta} = \Delta \frac{\boldsymbol{\chi}}{\|\boldsymbol{\chi}\|_2}$.
The parameter update rule becomes
\begin{equation}
    \boldsymbol{\theta}_{i+1} = \boldsymbol{\theta}_{i} -\eta \mathbf{G}(\boldsymbol{\theta})~.
\end{equation}

Note that in the standard SPSA~\cite{spallIntroduction2003}, each component $\Delta_\ell$ of the perturbation vector $\boldsymbol{\Delta}$ is typically drawn independently from a symmetric binary distribution, taking values $\pm 1$ with equal probability.
This results in perturbations restricted to one of $2^p$ discrete directions in parameter space. In contrast, our approach allows perturbations in arbitrary directions by sampling from a continuous uniform distribution, which we found leads to smoother behavior of the optimization and improved stability across iterations.

In both gradient descent and SPSA, using a fixed learning rate  has a drawback: if the learning rate is too small, convergence can be slow, requiring many iterations to reach the minimum; if it is too large, the updates may overshoot the minimum, especially when the parameters are close to optimal.  Although learning rate schedules --- where the learning rate is adjusted as a function of the iteration number --- can help mitigate these issues, the ADAM optimizer offers a significant improvement by adaptively tuning the learning rate for each individual variational parameter based on the cost function history.

This is achieved by introducing the ``momentum" $\mathbf{m}=(m_1,\cdots,m_p)$ and the ``velocity" $\mathbf{v}=(v_1,\cdots,v_p)$ of the parameters, both initialized to zero. These quantities are updated at each iteration according to the updating rules
\begin{align} \label{appeqn:adam momentum}
    \mathbf{m}_{i+1} &= \beta_1 \mathbf{m}_{i} + (1-\beta_1) \nabla C(\boldsymbol{\theta}_i) \\ 
    \label{appeqn:adam velocity}
    \mathbf{v}_{i+1} &= \beta_2 \mathbf{v}_{i} + (1-\beta_2) [\nabla C(\boldsymbol{\theta}_i)]^{\odot 2}~,
\end{align}
where $(\mathbf{v})^{\odot 2} := (v_1^2,v_2^2,\cdots,v_p^2)$ denotes element-wise squaring, and $(\beta_1,\beta_2) = (0.9, 0.999)$ are the hyperparameters.
To correct the initialization bias, one applies the following modifications
\begin{align}
    \tilde{\mathbf{m}}_{i+1} &=  \mathbf{m}_{i+1}/(1-\beta_1^{i+1}) \\
     \tilde{\mathbf{v}}_{i+1} &=  \mathbf{v}_{i+1}/(1-\beta_2^{i+1})~.
\end{align}
The parameters are then updated as 
\begin{equation}
    \boldsymbol{\theta}_{i+1} = \boldsymbol{\theta}_{i} - \eta \tilde{\mathbf{m}}_{i+1}/(\tilde{\mathbf{v}}_{i+1}^{\odot 0.5} + \epsilon)~,
\end{equation}
where $\mathbf{v}^{\odot 0.5} = (\sqrt{v_1},\cdots, \sqrt{v_p})$ is the element-wise square root, and all operations (division and addition) are element-wise.
The small hyperparameter $\epsilon = 10^{-5}$ is introduced to avoid division by zero.
To combine SPSA with ADAM, we replace $\nabla C(\boldsymbol{\theta})$ in Eqs.~(\ref{appeqn:adam momentum}) and (\ref{appeqn:adam velocity}) with the SPSA-estimated gradient  $\mathbf{G}(\boldsymbol{\theta})$ defined in Eq.~(\ref{appeqn: SPSA gradient}).

\section{Snake-ordering in DMRG}\label{app: snake-ordering DMRG}
\begin{figure*}
    \centering
    {\includegraphics[width=0.34\textwidth]{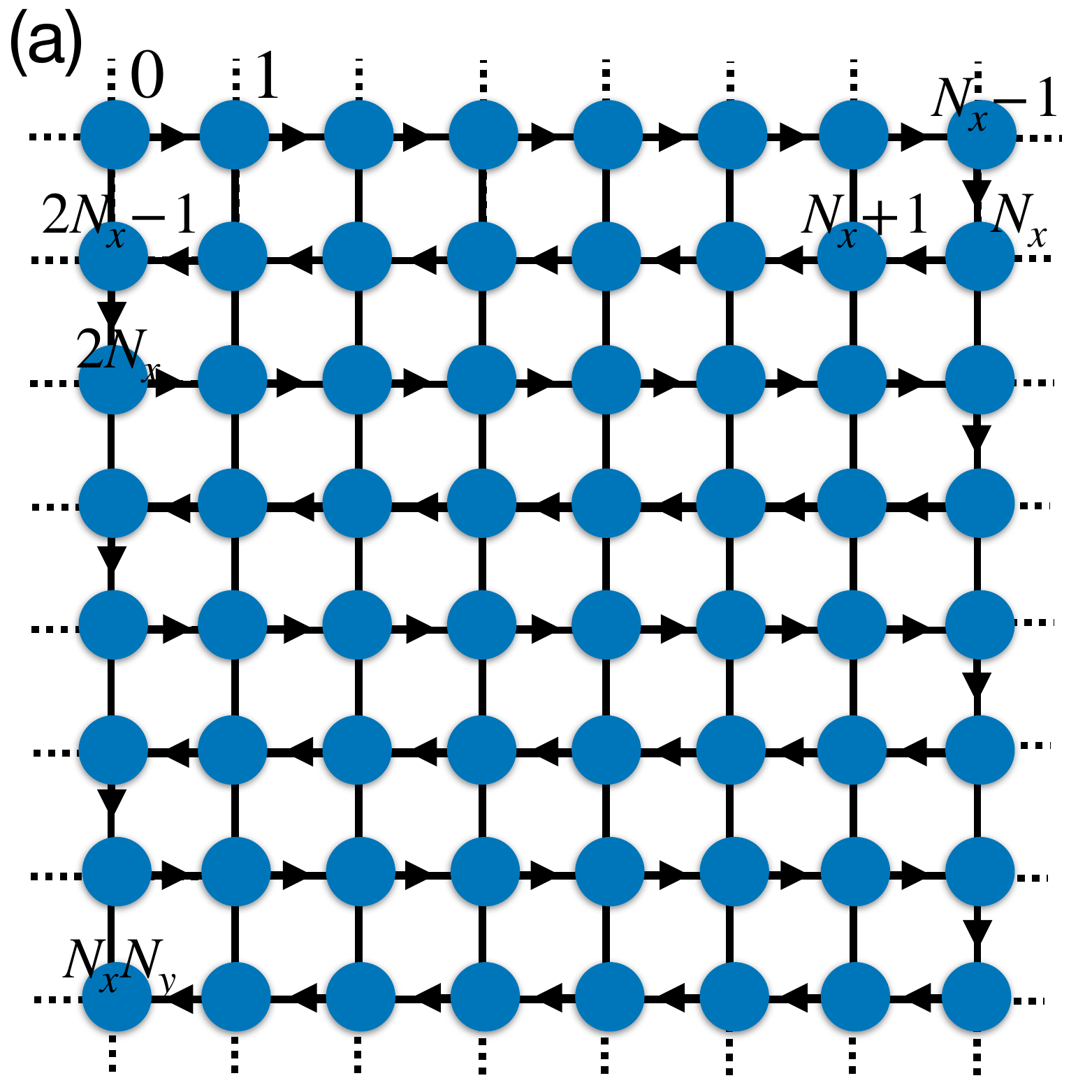}
    \includegraphics[width=0.45\textwidth]{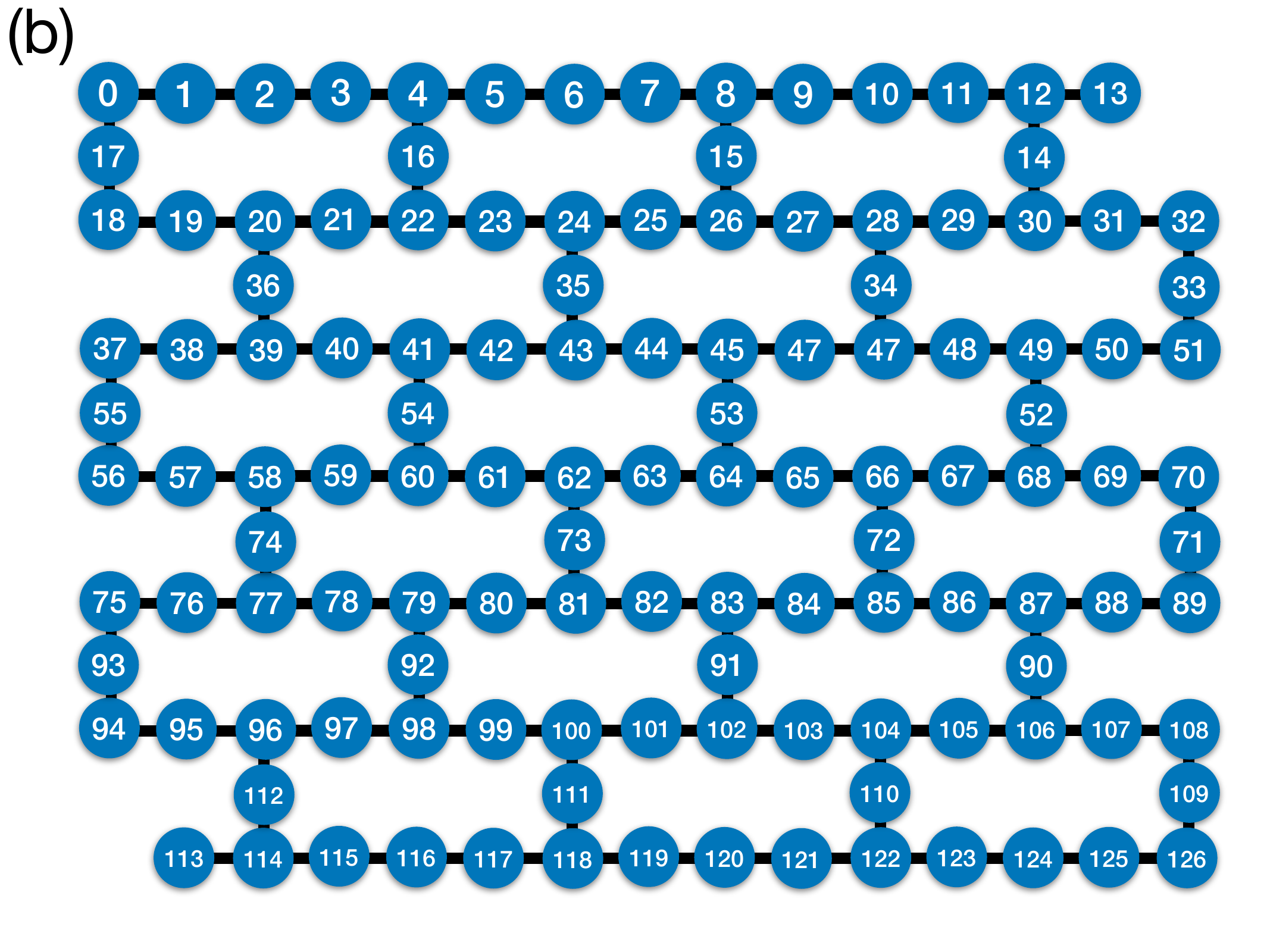}
    }
 
    \caption{\label{fig:snake order}The qubit ordering used in the DMRG calculation for (a) square lattice and (b) heavy hex lattice. The arrows indicate the ordering. For the square lattice, we consider the system size $N_x=N_y=10$, or $N=100$ qubits. For the heavy hex lattice, we have $N=127$ qubits.} 

\end{figure*}

In the DMRG calculations, we use snake-ordering to map the two-dimensional problem onto a one-dimensional chain with long-range interactions. The qubit orderings for the square lattice and the heavy-hex lattice are illustrated in Fig.~\ref{fig:snake order} (a) and (b), respectively.
For the square lattice with periodic boundary conditions, we consider a system of $N = 100$ qubits, while for the heavy-hex lattice with open boundary conditions, we use $N = 127$ qubits. In both cases, the bond dimension is set to $\chi = 500$.
While the DMRG trial energy could potentially be improved by increasing the bond dimension or by exploring alternative qubit orderings, we believe our results are representative of what non-experts can reasonably achieve using the ITensor library~\cite{itensor} without extensive DMRG tuning.

\section{Quantum circuit for flux-free state preparation}\label{app: fluxfree}
In this appendix, we describe the quantum circuit $V_{\text{f.f.}}$ used for to prepare the flux-free state. Our goal is to construct a state in which the plaquette operators satisfy $W_P=1$, and all $z$-type bonds satisfy $Z_iZ_j = +1$. This state corresponds to a ground state of the Kitaev honeycomb model with $(J_x, J_y, J_z) = (0, 0, 1)$ in the flux-free sector. 
Notably, the quantum circuit for this state can be derived from the ground-state preparation of the toric code, as we explain below.

\subsection{Mapping to a toric-code ground state}\label{appsub: map toric}

\begin{figure*}
    \centering
    \includegraphics[width=0.85\textwidth]{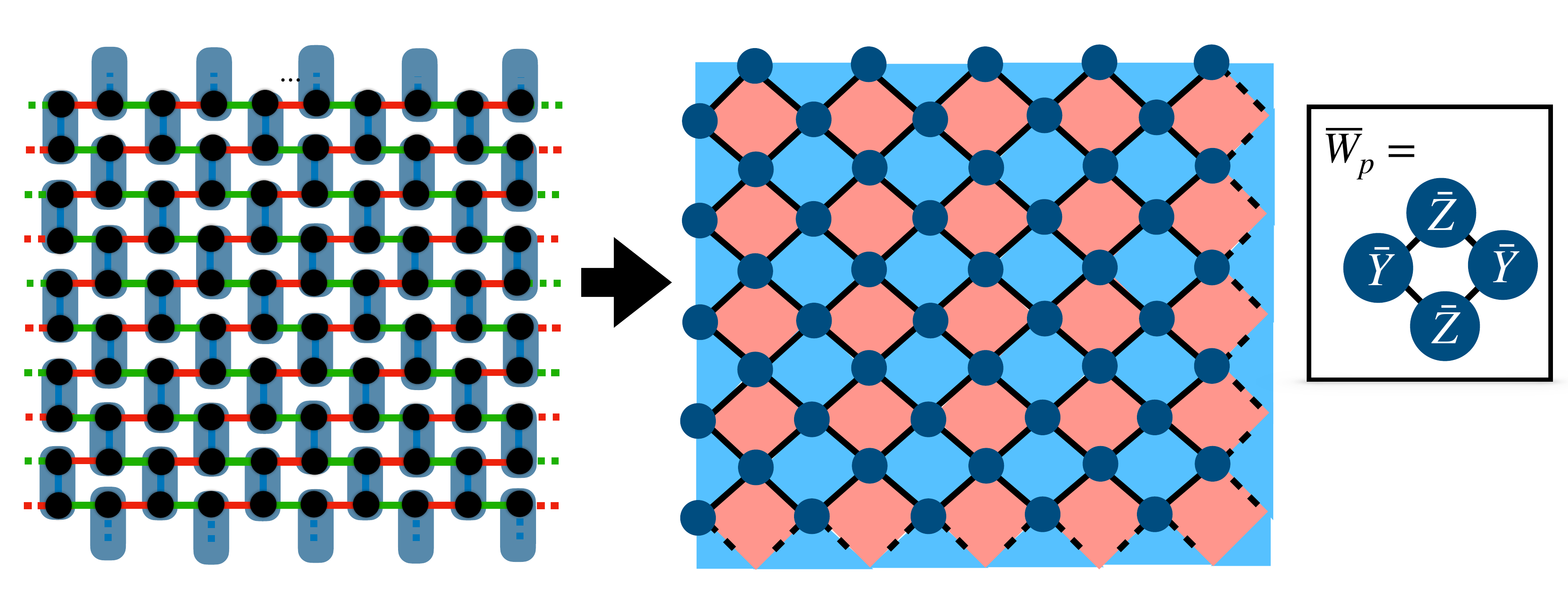}

    \caption{\label{fig:honeycomb to square} By projecting the qubits on the $z$-bonds $E_Z$ onto effective qubits ($|\bar{0}\ra = |00\ra$ and $|\bar{1}\ra = |11\ra$), the honeycomb lattice is mapped onto a rotated square lattice. The plaquette operator in the honeycomb model $W_P$ becomes the effective plaquette operator $\overline{W}_P$, as shown in the inset. They can be further transformed by the basis transformation $U_b^{\dagger}\overline{W}_PU_b = \tilde{W}_P$ to the more familiar toric-code stabilizers, where $\tilde{W}_P=\bar{X}\bar{X}\bar{X}\bar{X}$ for the red plaquettes and $\tilde{W}_P=\bar{Z}\bar{Z}\bar{Z}\bar{Z}$ for the blue plaquettes.}

\end{figure*}

\begin{figure*}
    \centering
    \includegraphics[width=0.85\textwidth]{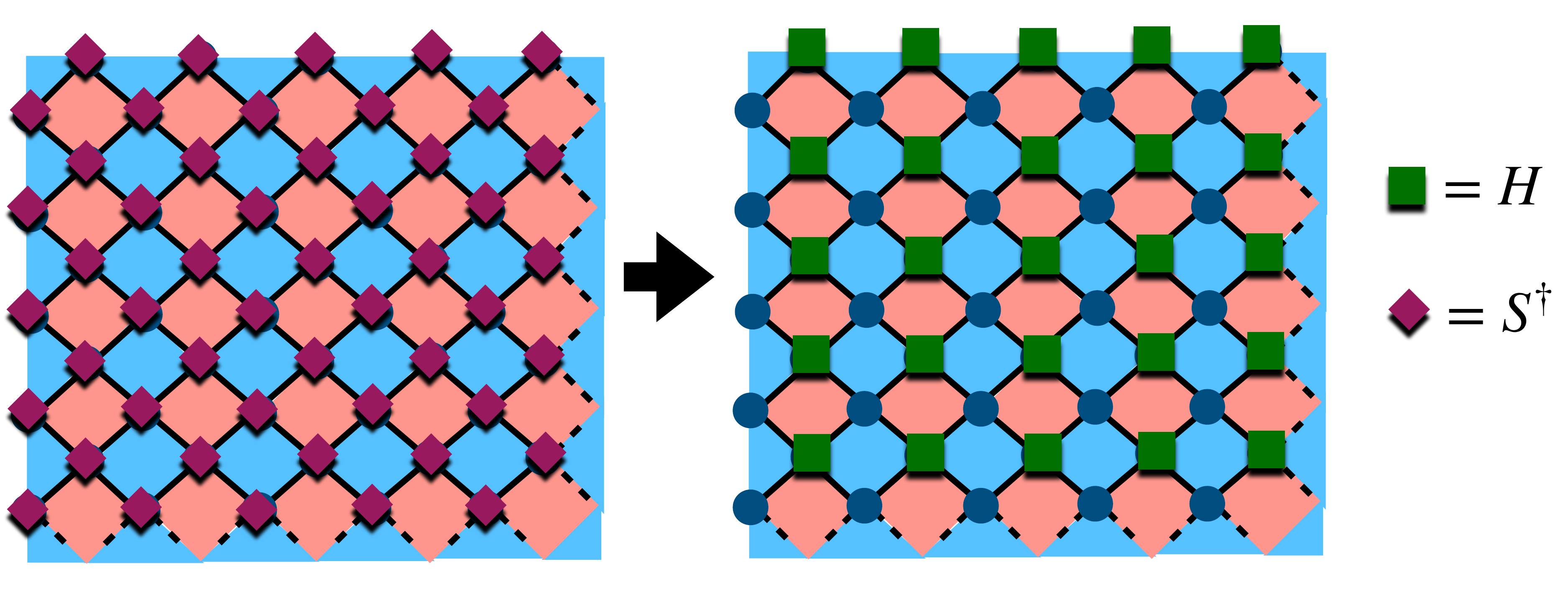}

    \caption{\label{fig:basis change unitary}  The basis transformation $U_b$ that maps the plaquette operators $\overline{W}_P$ into the more familiar toric-code plaquettes $\tilde{W}_P$. The circuit is operated by first applying the phase gates $S^{\dagger} = |0\rangle \langle 0| - i |1\rangle \langle 1|$ on all the qubits, and then applying the Hadamard gates on the qubits as indicated.
    } 

\end{figure*}

The mapping is achieved by treating  the two states satisfying $Z_iZ_j=+1$ on the $z$-bonds as an effective qubit. Specifically, we define $|\bar{0}\ra = |00\ra$ and $|\bar{1}\ra = |11\ra$. (We use the overline to denote the effective qubit and its operators. )
By replacing the $z$-type bonds on the honeycomb lattice with effective qubits, the system maps onto a ($45^{\circ}$-rotated) square lattice, as shown in Fig.~\ref{fig:honeycomb to square}. 
Projecting the plaquette operator $W_P$ into this effective-qubit subspace, we find $XY|00\ra = |11\ra $ and $XY|11\ra = |00\ra$, meaning that the operator $XY$ on the $z$-bonds corresponds to a $\bar{Y}$ operator on the effective qubit. It is also straightforward to verify  that the $Z$ operator in $W_P$ is an effective $\bar{Z}$ operator. Therefore, in terms of effective qubits, $\overline{W}_P=\bar{Y}\bar{Z}\bar{Y}\bar{Z}$ as depicted in Fig.~\ref{fig:honeycomb to square}.
These plaquette operators can be further transformed into a more familiar toric-code form via a unitary transformation $U_b$ on the effective qubits depicted in Fig.~\ref{fig:basis change unitary}, such that $\tilde{W}_P =U_b^{\dagger} \overline{W}_P U_b$~, where $\tilde{W}_P = \bar{X}\bar{X}\bar{X}\bar{X}$ for the red plaquettes and $\tilde{W}_P = \bar{Z}\bar{Z}\bar{Z}\bar{Z}$ for the blue plaquettes.

\subsection{Preparation of the Toric-code state on periodic boundary conditions}\label{appsub: preperation toric}

\begin{figure*}[!htbp]
    \centering
    \includegraphics[width=0.8\textwidth]{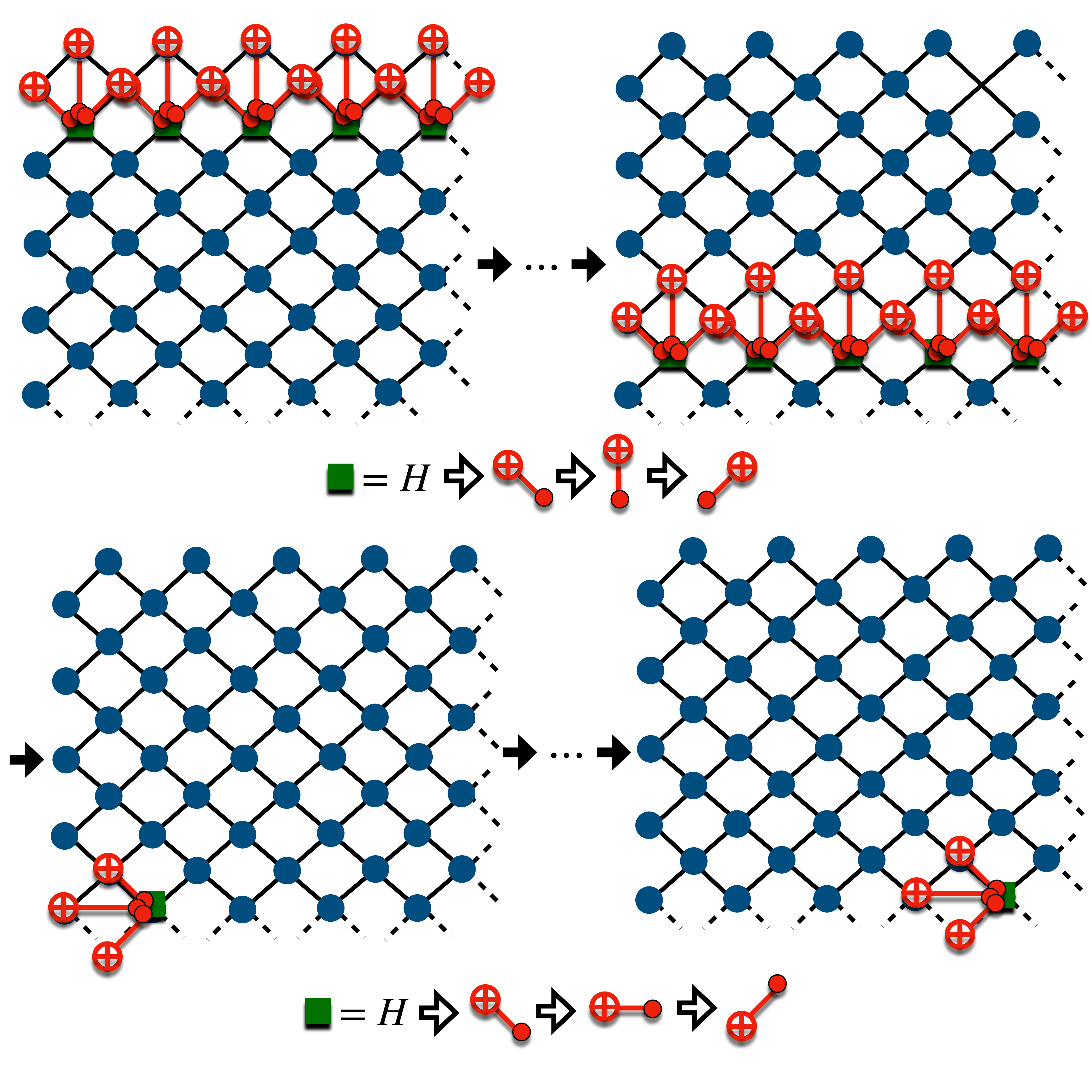}

    \caption{\label{fig:toric code unitary} The quantum circuit $U_{\text{toric}}$ prepares the toric-code ground state in which all plaquette operators satisfy $\tilde{W}_{P} = +1$. The operation proceeds row by row: starting from the first row to the second-to-last row, apply a Hadamard gate $H$ to the controlled qubits located at the bottom of each plaquette, followed by CNOT gates with the target qubits on the left, above, and then right.
    For the last row, the procedure runs from the left-most plaquette to the second-to-last plaquette. Apply a Hadamard gate to the right qubit of the plaquette, followed by CNOT gates with target qubits above, to the left, and then bottom.
    } 

\end{figure*}

To prepare the toric-code state in which all blue plaquettes satisfy $\tilde{W}_P = ZZZZ = +1$, and all red plaquettes satisfy $\tilde{W}_P = XXXX = +1$, we use the circuit $U_{\text{toric}}$ as shown in Fig.~\ref{fig:toric code unitary}. The circuit consists of Hadamard gates gates applied to the control qubits located at the bottom of each plaquette, followed by CNOT gates targeting the qubits to the left, above, and to the right of the plaquette.  This sequence is applied row by row, from the first row up to the second-to-last row.
For the last row, the procedure is applied from the leftmost plaquette to the second-to-last plaquette: a Hadamard gate is applied to the control qubit on the right of the plaquette, followed by CNOT gates targeting the qubits above, to the left, and below.

To understand why this circuit prepares the toric code state, it is helpful to analyze it in the Heisenberg picture, or equivalently, using the stabilizer formalism. 
The initial state $|0\ra^{\otimes n}$ is stabilized by the generators $Z_j = +1$. 
The Hadamard gate transforms the stabilizer according to $H Z H = X$, converting a $Z$ stabilizer into an $X$ stabilizer.
The CNOT gate “spreads” stabilizers as $\text{CNOT}_{12} X_1 \text{CNOT}_{12} = X_1 X_2$. Therefore, the action on a plaquette is to turn the control qubit’s stabilizer into $X$, while the CNOT gates extend this stabilizer into the four-qubit operator $XXXX$ on the corresponding plaquette.
The Z-type plaquettes are generated similarly through the ``spreading” effect of CNOT gates on Z operators, since $\text{CNOT}_{12} Z_2 \text{CNOT}_{12} = Z_1 Z_2$.

\subsection{Quantum gates on the effective qubits}\label{appsub: effective gates}
To translate the Toric-code state preparation circuit to the preparation of the flux-free state, we simply replace all gates in $U_{\text{toric}}$ with the corresponding gates acting on the effective qubit degrees of freedom.
These effective gates can be realized on the physical qubits as follows.

First, the effective Hadamard gate can be realized as
\begin{equation}
    \bar{H} = \text{CNOT}_{12}H_1\text{CNOT}_{12}~,
\end{equation}
which can be verified by its action on the effective qubit basis states
\begin{align}
    \bar{H}|\bar{0}\ra &= CNOT_{12}H_1CNOT_{12}|00\ra = \frac{1}{\sqrt{2}}(|00\ra + |11\ra) = \frac{1}{\sqrt{2}}(|\bar{0}\ra + |\bar{1}\ra) \notag \\
    \bar{H}|\bar{1}\ra &= CNOT_{12}H_1CNOT_{12}|11\ra = \frac{1}{\sqrt{2}}(|00\ra - |11\ra) = \frac{1}{\sqrt{2}}(|\bar{0}\ra - |\bar{1}\ra) \notag ~.
\end{align}
The effective $S$ gate is simply
\begin{equation}
    \bar{S}=S_1~.
\end{equation}
Finally, using the convention where qubits 1 and 2 form the first (control) effective qubit and qubits 3 and 4 form the second (target) effective qubit, the effective CNOT gate can be realized as 
\begin{equation}
    \text{CNOT}_{\bar{1}\bar{2}} = \text{CNOT}_{34}\text{CNOT}_{23}\text{CNOT}_{34}~.
\end{equation}

\subsection{Preparation of the flux-free state}
The preparation of the flux-free state is therefore given by $V_{\text{f.f}} = U_b^{\dagger} U_{\text{toric}}$, where $U_b$ and $U_{\text{toric}}$ are described in Appendices~\ref{appsub: map toric} and \ref{appsub: preperation toric} (or Figs.~\ref{fig:basis change unitary} and \ref{fig:toric code unitary}), repectively, with the gates replaced by the effective gates described in Appendix~\ref{appsub: effective gates}~.
Note that $V_{\text{f.f}}$ consists entirely of Clifford gates with  a circuit depth that scales linearly with the system size $N_x+N_y$.

Although not shown explicitly in this appendix, a similar strategy can be applied to prepare the flux-free state with open boundary conditions.
In this case, the problem is first mapped onto a toric-code ground state preparation on the effective qubit degrees of freedom with open boundary conditions, where some effective qubits on the boundary coincide with physical qubits. 
The toric-code ground state preparation circuit $U_{\text{toric}}$ can then be constructed from, say Ref.~\cite{satzingerRealizing2021}. After applying the basis transformation with the effective gates $U_b^{\dagger}$, one obtains the quantum circuit to construct flux-free states on open boundary conditions, such as the ones shown in Ref.~\cite{willProbing2025}.
\\

\twocolumngrid


%

\end{document}